\documentclass{aa}
\usepackage{graphicx}
\usepackage{color}
\usepackage[varg]{txfonts}
\graphicspath{{./Figures/}}
\defcitealias{Hyo21a}{Paper~I}
\def\Mearth{{\rm M}_\oplus}   
\def\Msun{{\rm M}_\odot}  

\begin{document}

\title{A ``no-drift'' runaway pile-up of pebbles in protoplanetary disks \\II. Characteristics of the resulting planetesimal belt}
\titlerunning{Planetesimal belt formed by the no-drift mechanism}

   \author{Ryuki Hyodo
          \inst{1}
          \and
          Shigeru Ida
          \inst{2}
	  \and
          Tristan Guillot
          \inst{3}
           }
         \institute{ISAS/JAXA, Sagamihara, Kanagawa, Japan (\email{hyodo@elsi.jp})
         \and Earth-Life Science Institute, Tokyo Institute of Technology, Meguro-ku, Tokyo 152-8550, Japan 
         \and Universit\'e C\^ote d'Azur, Laboratoire J.-L.\ Lagrange, CNRS, Observatoire de la C\^ote d'Azur, F-06304 Nice, France}
\date{DRAFT:  \today}

\abstract
{Forming planetesimals from pebbles is a major challenge in our current understanding of planet formation. In a protoplanetary disk, pebbles drift inward near the disk midplane via gas drag and they may enter a zone of reduced turbulence (dead zone). In this context, we identified that the backreaction of the drag of pebbles onto the gas could lead to a runaway pile-up of pebbles, the so-called no-drift mechanism for the formation of planetesimals.}
{We improve upon the previous study of planetesimal formation from accumulating pebbles via the no-drift mechanism by investigating the nature and characteristics of the resultant planetesimal belt.} 
{We performed 1D diffusion-advection simulations of drifting pebbles in the outer region of a modeled dead zone by including a pebble-gas backreaction to the radial drift of pebbles and including planetesimal formation via the streaming instability. We independently considered the parameters that regulate gas accretion ($\alpha_{\rm acc}$) and vertical stirring of pebbles in the disk midplane ($\alpha_{\rm mid}$). In this study, the pebble-to-gas mass flux ($F_{\rm p/g}$) was fixed as a parameter.}
{We find that, for a constant $F_{\rm p/g}$, after the criteria of the no-drift mechanism are satisfied, planetesimals initially form within a narrow ring whose width expands as accumulating pebbles radially diffuse over time. The system finally reaches a steady-state where the width of the planetesimal belt no longer changes, followed by a continuous formation of planetesimals. A non-negligible total mass of planetesimals (more than one Earth mass) is formed via the no-drift mechanism for a disk having $F_{\rm p/g} \gtrsim 0.1$ for more than $\sim 10-100$\,kyr with nominal parameters: a gas mass flux of $\gtrsim 10^{-8}\,{\Msun}$\,yr$^{-1}$, $\tau_{\rm s} \simeq 0.01-0.1$, $\alpha_{\rm mid} \lesssim 10^{-4}$, and $\alpha_{\rm acc} \simeq 10^{-3}-10^{-2}$ at $r\lesssim10$\,au, where $r$, $\tau_{\rm s}$, $\alpha_{\rm mid}$, and $\alpha_{\rm acc}$ are the heliocentric distance, the Stokes number, and the parameters in a dead zone controlling the efficiencies of vertical turbulent diffusion of pebbles (i.e., scale height of pebbles) and gas accretion of the $\alpha$-disk (i.e., gas surface density), respectively.} 
 {}

\authorrunning{R. Hyodo, S. Ida, T. Guillot}
\maketitle 

\section{Introduction} \label{sec_intro}
Planetesimals, measuring a subkilometer to several hundred kilometers, are fundamental building blocks of planets and small bodies \citep{Saf72,Hay85}. It is thought that planetesimals form from micron-sized dust in protoplanetary disks. 

However, forming planetesimals from small dust through all intermediate sizes via pairwise collisions is still challenging. Two theoretical challenges exist -- the growth barrier and the radial drift barrier. The growth barrier is caused by the fact that approximately centimeter-sized icy particles become too large to grow further due to fragmentation or bouncing during their high-speed collisions \citep{Blu00,Zso10}. The radial drift barrier arises because the radial drift due to gas drag becomes too fast for meter-sized particles to grow to planetesimals before they fall onto the host star \citep{Whi72,Wei77}.

An alternative idea is gravitational collapse for the planetesimal formation from small particles (typically micron-sized dust and/or millimeter- to decimeter-sized pebbles). A gravitational instability \citep[hereafter GI;][]{Gol73,You02,Ger20} and/or the streaming instability \citep[hereafter SI;][]{You05,Joh07a} are possible mechanisms that can avoid the above challenges and that can lead to planetesimal formation without a step-by-step gradual growth, although a high spatial particle concentration within the gas disk is required for triggering these mechanisms \citep{Car15,Yan17,You02,Ger20}. 

In this study, a follow-up study of the no-drift (hereafter ND) mechanism was performed. The ND mechanism is a newly reported physical mechanism that causes a traffic jam of pebbles and that leads to a runaway accumulation of pebbles in disks, thus favoring the formation of planetesimals by SI or GI \citep[][hereafter \citetalias{Hyo21a}]{Hyo21a}. The ND mechanism differs from other mechanisms in that it does not require a pressure bump, snow line, and pebble growth and/or fragmentation.

The main drivers of the ND mechanism are (1) a pebble-gas friction, backreaction, that slows down the radial drift of pebbles as its spatial concentration to that of the gas increases, and (2) different $\alpha$-parameters for the gas accretion, $\alpha_{\rm acc}$, and for the midplane turbulence, $\alpha_{\rm mid}$, that regulates the pebble vertical scale height. Such a condition should occur when the gas accretion is regulated by a different mechanism from that for the vertical turbulence near the disk midplane -- for example, a layered disk with a disk wind could trigger the gas accretion only near the disk surface \citep[e.g.,][]{Arm11,Bai13,Lee21}. 

\citetalias{Hyo21a} analytically and numerically investigated the aerodynamically coupled 1D system of drifting pebbles and gas within an outer part of a dead zone, assuming the gas profile is described by the classical $\alpha$-accretion disk model \citep[i.e., the gas surface density $\Sigma_{\rm g} \propto \alpha_{\rm acc}^{-1}$;][]{Sha73,Lyn74} and the vertical turbulent $\alpha_{\rm mid}$ decreases with decreasing $r$ (i.e., the outer region of a dead zone). For a large $\alpha_{\rm acc}$, $\Sigma_{\rm g}$ decreases and correspondingly the midplane pebble-to-gas ratio increases. For a small $\alpha_{\rm mid}$, the scale height of pebbles decreases and correspondingly the radial backreaction more efficiently slows down the radial drift of pebbles as the midplane pebble-to-gas ratio increases. 

It showed that a runaway accumulation of pebbles (i.e., no-drift), positive feedback, occurs when the pebble-to-gas mass flux, $F_{\rm p/g}$, is large enough and when drifting pebbles entering a dead zone reach a critical level of low-turbulence (i.e., a critical small $\alpha_{\rm mid}/\alpha_{\rm acc}$ which depends on $F_{\rm p/g}$). This occurs because of the traffic jam effect -- that is, at large orbital distances (corresponding to a large $\alpha_{\rm mid}$), pebbles drift inward efficiently, whereas the pebble drift is progressively seized in the dead zone (where $\alpha_{\rm mid}$ becomes small) via the backreaction effect, leading to a runaway accumulation of pebbles.

\citetalias{Hyo21a}, however, did not include the physics of planetesimal formation after the point where pebbles accumulate in a runaway fashion due to the ND mechanism, that is, pebble surface density and local spatial density simply kept increasing (see their Figure 2). In reality, GI or SI would then operate to form planetesimals. The resultant planetesimal distribution should be characterized by the balance between the rate of pebble pile-up by the ND mode and that of planetesimal formation from pebbles by GI or SI\footnote{For pebbles, SI is generally favored over GI as its required solid-to-gas ratio is smaller. Thus, in this study, we considered SI as the main driver of the formation of planetesimals from accumulating pebbles.}.

In this study, an analytical algorithm that forms planetesimals from accumulating pebbles via SI was implemented in our 1D numerical code to further study the ND mechanism. We address (1) the resultant planetesimal belt, (2) its dependencies on different physical parameters, and (3) the rate of planetesimal formation by the ND mode.

The structure of this paper is as follows: in Sect.~\ref{sec_methods}, we introduce concepts and formulae of our numerical approach. Section \ref{sec_results} presents some fiducial simulation results and discusses its dependencies. Section \ref{sec_discussion} discusses possible applications of the ND mechanism. Our findings are summarized in Sect.~\ref{sec_summary}.

\section{Concepts and formulae of our numerical approach} \label{sec_methods}
The numerical approach is the same as \citetalias{Hyo21a} except for additional consideration of planetesimal formation from accumulating pebbles via SI. Below, we briefly explain the concepts and formulae of our numerical approach.

\subsection{Disk structure}
A 1D radial protoplanetary disk, a function of the distance to the central star $r$, was constructed using the classical $\alpha$-accretion disk model \citep{Sha73,Lyn74}. The surface density of gas in a steady accretion disk is given by
\begin{equation}
	\Sigma_{\rm g} = \frac{\dot{M}_{\rm g}}{3\pi \nu_{\rm acc}} =  \frac{\dot{M}_{\rm g}}{3\pi \alpha_{\rm acc} c_{\rm s}^{2} \Omega_{\rm K}^{-1}},
\label{eq_sigma_g}
\end{equation}
where $\dot{M}_{\rm g}$ and $\nu_{\rm acc}=\alpha_{\rm acc} c_{\rm s}^{2} \Omega_{\rm K}^{-1}$ are the gas mass accretion rate and the effective viscosity, respectively ($c_{\rm s}$ being the gas sound velocity and $\Omega_{\rm K}$ being the Keplerian orbital frequency).

The disk radial structure is characterized by its radial pressure gradient using $C_{\rm \eta}$ as
\begin{equation}
	C_{\rm \eta} \equiv -\frac{1}{2} \frac{\partial \ln P_{\rm g}}{\partial \ln r},
\end{equation}
where $P_{\rm g}$ is the gas pressure which depends on the disk temperature profile, $T(r)$ (e.g., $C_{\rm \eta} = 11/8$ for $T \propto r^{-1/2}$).

In this study, we considered a dead zone in the disk midplane (Sect.~\ref{sec_turbulence}) and the above smooth gas disk structure was fixed throughout the simulations assuming that the efficient gas accretion is driven by, for example, disk winds in the upper layers. In reality, a coupled simulation including protoplanetary disk evolution is required to address the fate of a specific system (Sect.~\ref{sec_discussion}).

 \begin{table}[t]
\caption{Parameters used to model the dead zone (Eq.~(\ref{eq_alpha_mid})).}
\begin{tabular}{cllll}  \hline
Disk models  & $r^{*}$ & $\Delta r_{\rm tra}$ & $\alpha_{\rm acc}$ & $\alpha_{\rm dead}$  \\ \hline \hline
 Disk 1  & 5 au   & 0.5 au   & $10^{-2}$ & $10^{-4}$ \\ \hline
 Disk 2  & 9 au   & 5.0 au   & $10^{-2} - 10^{-3}$ & $10^{-4}$ \\ \hline 
 Disk 3  & 20 au & 18.0 au & $10^{-2}$ & $10^{-4}$ \\ \hline 
\hline
\end{tabular}
\label{table_disk}
\end{table}

\subsection{Pebble radial drift}

The driver of the ND mechanism is the radial backreaction, inertia, that slows down the radial velocity of pebbles for high values of the midplane pebble-to-gas ratio. Including backreaction, the radial velocity of pebbles, $v_{\rm p}$, is given as \citep{Ida16,Sch17}
\begin{align}
\label{eq_vp}
	v_{\rm p} = - \frac{\Lambda}{1+\Lambda^{2}\tau_{\rm s}^{2}} \left( 2\tau_{\rm s}\Lambda \eta v_{\rm K} - v_{\rm g} \right) ,
\end{align}
where  $\Lambda \equiv \rho_{\rm g}/(\rho_{\rm g} + \rho_{\rm p}) = 1/(1+Z)$ characterizes the strength of the backreaction due to the pile-up of pebbles ($Z \equiv \rho_{\rm p}/\rho_{\rm g}$ being the midplane pebble-to-gas spatial density ratio). Here, a negative value indicates radial drift to the central star. $\rho_{\rm g} \equiv \Sigma_{\rm g}/\sqrt{2\pi}H_{\rm g}$ and $\rho_{\rm p} \equiv \Sigma_{\rm p}/\sqrt{2\pi} H_{\rm p}$ where $\Sigma_{\rm g}$,  $\Sigma_{\rm p}$, $H_{\rm g}$, and $H_{\rm p}$ being the surface densities and scale heights of gas and pebbles, respectively. $v_{\rm K}$ and $\eta \equiv C_{\rm \eta} (H_{\rm g}/r)^2$ are the Keplerian velocity and the degree of deviation of the gas rotation frequency from that of Keplerian, respectively. $v_{\rm g} = -3 \nu_{\rm acc}/(2r)$ is the gas accretion velocity.

In this study, pebbles were considered to be decoupled from the gas with a Stokes number $\tau_{\rm s}=0.01 - 0.1$ \citep[e.g.,][]{Oku12,Ida16b}. Changing $\tau_{\rm s}$ linearly changes the critical value of $F_{\rm p/g}$ for the ND mode to be triggered \citepalias[Eq.(16) in][see also Eq.~(\ref{eq_ND_cri})]{Hyo21a}.

\subsection{Scale height of pebbles}
Another essential feature of the ND process is the midplane pebbles-to-gas concentration of the spatial density (i.e., $Z \equiv \rho_{\rm p}/\rho_{\rm g}$) that directly affects $\Lambda$ through $Z$ (Eq. \ref{eq_vp}). As $\rho_{\rm p} \propto 1/H_{\rm p}$, a  small scale height of pebbles elevates the midplane pebble-to-gas density ratio, leading to a less efficient radial drift of pebbles via the radial backreaction. In this study, for $H_{\rm p}$, the larger of the two -- $H_{\rm p,tur}$ and $H_{\rm p,KH}$ -- was adopted as
\begin{align}
\label{eq_Hp}
	H_{\rm p} = \max \left\{ H_{\rm p,tur}, H_{\rm p,KH} \right\} ,
\end{align}
where $H_{\rm p,tur}$ is regulated by the midplane vertical turbulence and $H_{\rm p,KH}$ is regulated by a vertical shear Kelvin-Helmholtz (KH) instability.

When the midplane vertical turbulence ($\propto \alpha_{\rm mid}$) regulates the vertical scale height of pebbles, it is given as \citep{Hyo19}
\begin{align}
\label{eq_Hp_tur}
	H_{\rm p,tur} = \left( 1 + \frac{\tau_{\rm s}}{\alpha_{\rm mid} \left( 1+ Z \right)^{-K} } \right)^{-1/2} H_{\rm g} ,
\end{align}
where a coefficient $K$ characterizes the strength of the backreaction onto the diffusivity and $K=1$ is used.
 
When the turbulent $\alpha$-parameter, $\alpha_{\rm mid}$, is very small, a KH instability may prevent a further decrease in the pebble scale height. Thus, the minimum scale height, $H_{\rm p, KH}$ ($Ri=0.5$), is defined as 
\begin{align}
	H_{\rm p,KH} \simeq Ri^{1/2} \frac{Z^{1/2}}{ \left( 1+Z \right)^{3/2} } C_{\rm \eta} \left( \frac{H_{\rm g}}{r} \right) H_{\rm g} = Ri^{1/2} \frac{Z^{1/2}}{ \left( 1+Z \right)^{3/2} } \eta r,
\label{eq_Hp_KH}
\end{align}
which is generally valid for $Z \lesssim 1$, and $Z \gg 1$ indicates a gravitational collapse \citep{Hyo21b}. We note that \cite{Chi08} solved the particle height distribution function and derived $z_{\rm max}$ where the particle distribution is truncated. Their distribution function has a cusp for $Z  \gtrsim 1$ and the cusp becomes sharper as $Z$ increases. Our definition of $H_{\rm p,KH}$ is based on the root mean square of the heights of the particle distribution. The difference between $z_{\rm max}$ of \cite{Chi08} and our $H_{\rm p,KH}$ becomes larger as $Z$ increases, while they are similar (only differ by $2$) for $Z \ll 1$ \citep[see more details in][]{Hyo21b}. For $\alpha_{\rm mid}=10^{-4}$, $H_{\rm p,tur}>H_{\rm p,KH}$ and thus $H_{\rm p}=H_{\rm tur}$. For $\alpha_{\rm mid}<10^{-4}$, $H_{\rm p}=H_{\rm p,KH}$ depending on $Z$ during the evolution (Fig.~\ref{fig_parameter}). Here, to be consistent with the settings in \citetalias{Hyo21a} and Fig.~\ref{fig_parameter}, we used Eq.~(\ref{eq_Hp}) in our numerical simulations. We used $\alpha_{\rm mid} \ge 10^{-4}$ and thus our choise of $H_{\rm p,KH}$ does not affect our numerical results.

\subsection{Midplane turbulent structure}
\label{sec_turbulence}
In our numerical approach, we used a nondimensional midplane turbulent $\alpha$-parameter, $\alpha_{\rm mid}$, which differs from $\alpha$-parameter that characterizes the gas accretion, $\alpha_{\rm acc}$ (Eq.~(\ref{eq_sigma_g})). We adopted a dead zone in the inner region of the disk midplane modeled as
\begin{equation}
	\alpha_{\rm mid}(r) = \alpha_{\rm dead} +  \left( \frac{\alpha_{\rm acc} - \alpha_{\rm dead}}{2} \right) \left[ \mathrm{erf} \left( 3 + \frac{6\left( r - r^{*} \right) }{\Delta r_{\rm tra}} \right) + 1 \right],
\label{eq_alpha_mid}
\end{equation}
where $\alpha_{\rm acc}$ and $\alpha_{\rm dead}$ are turbulence parameters outside and inside the dead zone. $r^{*}$ is the innermost radial distance where $\alpha_{\rm mid}=\alpha_{\rm acc}$. $\Delta r_{\rm tra}$ is the radial width of the transition from $\alpha_{\rm mid}=\alpha_{\rm acc}$ to $\alpha_{\rm mid}=\alpha_{\rm dead}$. Although it is still under intense research and depends on the detailed settings of the hydrodynamics simulations, $\alpha_{\rm acc}$ could be up to $\alpha_{\rm acc} \sim 10^{-2}$ in well-ionized gas \citep[e.g.,][]{Flo17,Flo19}, while $\alpha_{\rm dead} \sim 10^{-5}-10^{-3}$ \citep[e.g.,][]{Oku11,Mor19}. The radial width of the transition from an active zone to a dead zone (i.e., $\Delta r_{\rm tra}$) could be $\gtrsim 10$\,au \citep[e.g.,][]{Dzy13} or as small as the local scale height ($\sim 0.1$\,au; Okuzumi personal communication). We used $r^{*}$, $\Delta r_{\rm tra}$, $\alpha_{\rm acc}$, and $\alpha_{\rm dead}$ as parameters in this study (Table \ref{table_disk} and see \cite{Pin16} for similar settings; Fig.~\ref{fig_planetesmial} top panels). Our chosen parameters triggered the ND mechanism at $r \lesssim 10$\,au for $F_{\rm p/g} \gtrsim 0.1$.

As shown in \citetalias{Hyo21a}, the ND mode occurs for an arbitrary choice of the dead zone's radial structure, that is, irrespective of a sharp or a smooth change between the active and dead zones (i.e., arbitrary choice of $\alpha_{\rm dead}$, $\alpha_{\rm acc}$, $r^{*}$, and $\Delta r_{\rm tra}$) as long as $\alpha_{\rm mid}$ is smaller than a threshold value (Eq.~(\ref{eq_ND_cri}); derived in \citetalias{Hyo21a}).

\subsection{From pebbles to planetesimals}
The above settings and models are the same as those in \citetalias{Hyo21a}. In this follow-up study, we additionally included a toy model of planetesimal formation via SI in which a fraction of pebbles are converted to planetesimals. 

When $Z \geq 1$ and $\tau_{\rm s} \geq 0.01$ are locally satisfied, pebbles lead to the formation of planetesimals as \citep[][]{Dra16}
\begin{align}
	\frac{ d \Sigma_{\rm pla}}{d t} &= k_{\rm SI} \Sigma_{\rm p} \Omega \nonumber \\ 
	\frac{ d \Sigma_{\rm p}}{d t} &= -k_{\rm SI} \Sigma_{\rm p} \Omega ,
\label{eq_planetesimal}
\end{align}
where $\Sigma_{\rm pla}$ is the surface density of planetesimals. $k_{\rm SI}$ is a coefficient that characterizes the efficiency of conversion from pebbles to planetesimals. It can be rewritten as $k_{\rm SI}=\zeta \tau_{\rm s}$ (a coefficient $\zeta=0.1$ was used in e.g., \cite{Sta19,Mil21}). Thus, for $\tau_{\rm s}=0.1$, $\zeta=0.1$ corresponds to $k_{\rm SI}=10^{-2}$. We used $k_{\rm SI}=10^{-2} - 10^{-4}$.

\subsection{Numerical settings and parameters}
We performed 1D diffusion-advection simulations that included the backreaction to radial drift of pebbles that slows down the pebble drift for an elevated midplane pebble-to-gas concentration. For the gas accretion, $\alpha_{\rm acc}=1 \times 10^{-3}$, $3 \times 10^{-3}$, and $1 \times10^{-2}$ were used. The governing equation of pebbles is given as \citep{Des17}
\begin{align}
\label{eq_sigma_peb}
	 \frac{\partial \Sigma_{\rm p}}{\partial t} =  -\frac{1}{r}\frac{\partial}{\partial r} \left( r \Sigma_{\rm p} v_{\rm p} - r D_{\rm p}\Sigma_{\rm g} \frac{\partial}{\partial r} \left( \frac{\Sigma_{\rm p}}{\Sigma_{\rm g} } \right) \right) ,
\end{align}
where $D_{\rm p} = \alpha_{\rm mid}c_{\rm s}^{2} \Omega_{\rm K}^{-1} \Lambda^{K} / (1+{\rm \tau_{\rm s}^2})$ is the radial diffusivity of pebbles. Here, the diffusivity included the effects of the backreaction with $K=1$ \citep{Hyo21a}. The temperature profile was $T(r) = 150 {\, \rm K} \times ( r/3 {\, \rm au} )^{-1/2}$. This leads to $C_{\rm \eta} = 11/8$ and $\Sigma_{\rm g} \propto r^{-1}$. The central star had the mass of the Sun, $M_{\odot}$. The gas molecular weight was $\mu_{\rm g}=2.34$. $\dot{M}_{\rm g}=10^{-9}, 10^{-8}$, and $10^{-7}\,M_{\odot}$\,yr$^{-1}$ were adopted. 
  
At the beginning of the 1D simulations, we set the pebble-to-gas mass flux $F_{\rm p/g} \equiv \dot{M}_{\rm p}/\dot{M}_{\rm g}$ (where $\dot{M}_{\rm p}$ is the pebble mass flux) at the outer boundary, $r_{\rm out}$ (typically $r_{\rm out}=30$ au). $F_{\rm p/g}$ at $r_{\rm out}$ was fixed throughout the simulations. We set $F_{\rm p/g} \sim 0.1 - 0.5$ as a typical value. With this value of $F_{\rm p/g}$, $\dot{M}_{\rm p} \sim 10^{-4} -10^{-3} {\Mearth}$ yr$^{-1}$ corresponds to $\dot{M}_{\rm g} \sim 10^{-9} - 10^{-7} M_{\odot}$ yr$^{-1}$, where $\Mearth$ is the Earth-mass. This is consistent with \citet{Dra21} (see their case with $\alpha_{\rm mid} = 10^{-4}$ and including pebble fragmentation). When pebble fragmentation is not effective, $F_{\rm p/g}$ could become about the order of unity for $\alpha_{\rm mid} \sim 10^{-4}$ and $\dot{M}_{\rm g} \sim 10^{-8}\,M_{\odot}$\,yr$^{-1}$ \citep{Ida16b}.

In this study, planetesimals do not interact with disks and pebbles, and the growth of planetesimals is not considered. The pebble/planetesimal-to-planet formation \citep[e.g.,][]{Orm10,Lam14,Liu19,Lic21} in the context of the ND mechanism should be studied in the future.

\section{Results} \label{sec_results}

Below, we show our numerical results as well as analytical estimates. First, we present overall parameters in the $\alpha_{\rm mid}/\alpha_{\rm acc} - F_{\rm p/g}$ space where the ND mechanism is triggered (Sect.~\ref{sec_parameter}). Second, including planetesimal formation, we introduce overall numerical results which demonstrate that the resultant planetesimal belts formed by the ND mechanism could be diverse in terms of their widths and profiles (Sect.~{\ref{sec_overall}}). Third, we show the dependencies on different parameters (e.g., $k_{\rm SI}$ and $\dot{M}_{\rm g}$; Sects.~{\ref{sec_kSI} and \ref{sec_dotMg}). Forth, we discuss the inner and outer edges of the planetesimal belts (Sect.~{\ref{sec_edge}}). Fifth, the rate of planetesimal formation via the ND mode is presented (Sect.~{\ref{sec_mass}}). Sixth, the dependence on the Stokes number, $\tau_{\rm s}$, is discussed (Sect.~{\ref{sec_tau}}). Finally, the dependence on $\alpha_{\rm acc}$ is discussed (Sect.~{\ref{sec_acc}}).

\subsection{Parameters triggering the ND mechanism} \label{sec_parameter}

\begin{figure*}[t]
	\centering
	\resizebox{\hsize}{!}{ \includegraphics{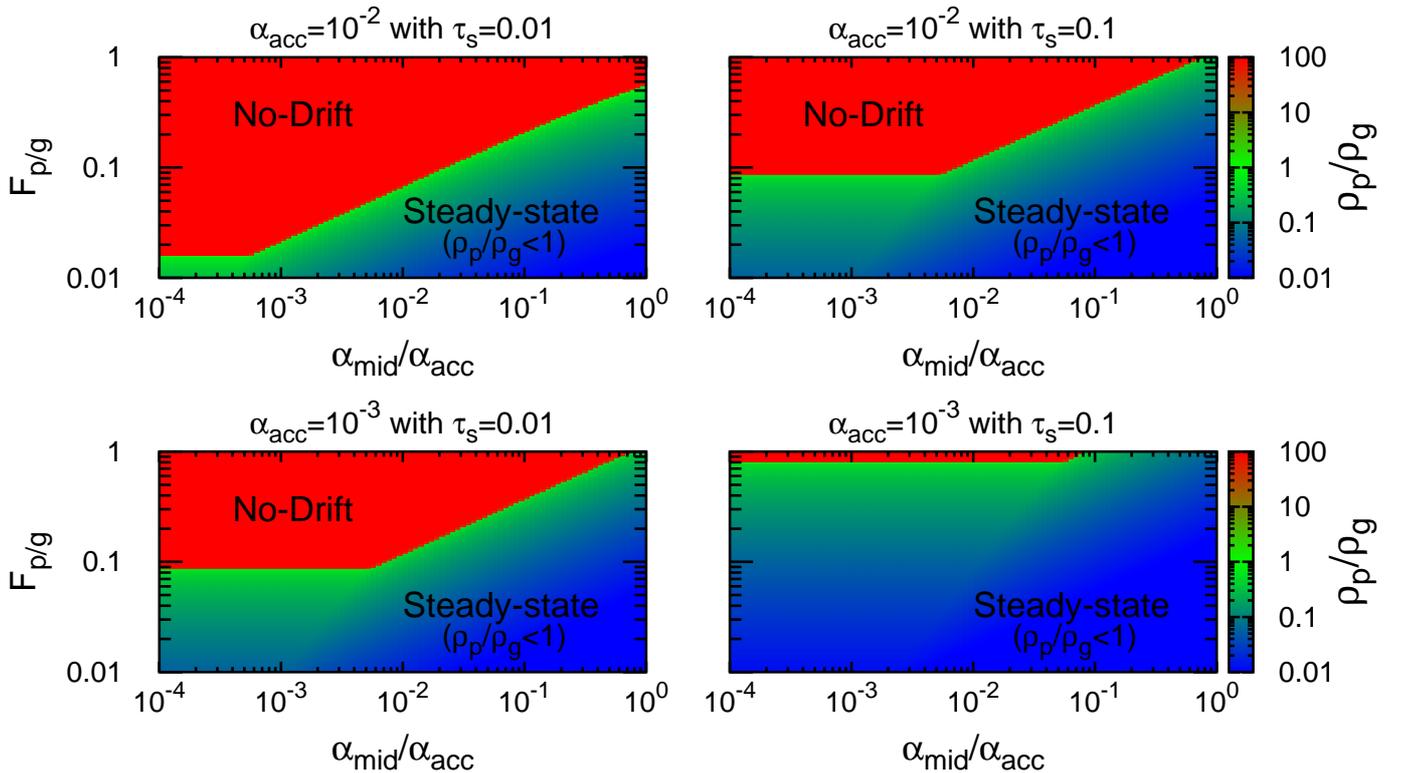} }
	\caption{Midplane pebble-to-gas ratio $Z \equiv \rho_{\rm p}/\rho_{\rm g}$ in the $\alpha_{\rm mid}/\alpha_{\rm acc} - F_{\rm p/g}$ space (Eq.~(\ref{eq_peb_pileup})). Top and bottom panels correspond to $\alpha_{\rm acc}=10^{-2}$ and $\alpha_{\rm acc}=10^{-3}$, respectively. Left and right panels correspond to $\tau_{\rm s}=0.01$ and $\tau_{\rm s}=0.1$, respectively. The color contours are obtained by directly solving Eq.~(\ref{eq_peb_pileup}). The red-colored regions indicate the "no-drift" runaway pile-up (i.e., $\rho_{\rm p}/\rho_{\rm g}$ goes to infinity; no-solution). The calculations correspond to $r=5$\,au, but the results depend very weakly on the radial distance to the star. The ND mode can occur irrespectively of the radial shape and position of a dead zone (see more details in \citetalias{Hyo21a}). We note that the physics of planetesimal formation is not included here.}
\label{fig_parameter}
\end{figure*}

Here, before performing the 1D numerical simulations that include planetesimal formation, we first present the overall parameters in the $\alpha_{\rm mid}/\alpha_{\rm acc} - F_{\rm p/g}$ space where the ND mechanism is triggered (Fig.~\ref{fig_parameter}). The arguments are made by directly solving Eq.~(\ref{eq_peb_pileup}) for $\alpha_{\rm acc}=10^{-3}$ and $10^{-2}$ with $\tau_{\rm s}=0.01$ and $0.1$ to find parameter sets for the ND mode to be triggered. This is originally demonstrated in \citetalias{Hyo21a}, although it was limited to $\tau_{\rm s}=0.1$.

The concentration of pebbles at the midplane is written as
\begin{equation}
	Z \equiv \frac{\rho_{\rm p}}{\rho_{\rm g}} = \frac{\Sigma_{\rm p}}{\Sigma_{\rm g}} h_{\rm p/g}^{-1} = \frac{v_{\rm g}}{v_{\rm p}} h_{\rm p/g}^{-1} F_{\rm p/g} ,
\label{eq_peb_pileup}
\end{equation}
where $h_{\rm p/g} \equiv H_{\rm p}/H_{\rm g}$. Eq.~(\ref{eq_peb_pileup}) indicates the existence of a positive feedback. When $v_{\rm p}$ decreases, $Z$ increases and then $v_{\rm p}$ further decreases due to the radial backreaction (Eq.~\ref{eq_vp}). This feedback can be triggered when $h_{\rm p/g}$ decreases (i.e., $Z$ increases and thus $v_{\rm p}$ decreases) as pebbles enter a dead zone.

Now, Eq.~(\ref{eq_peb_pileup}) can be solved (i.e., solve for $Z$) for a given combination of $F_{\rm p/g}$ and $\alpha_{\rm mid}$. The color contours in Fig.~\ref{fig_parameter} show the values of $Z$ obtained in a steady-state with a red-shaded region indicating that no steady-state solution is found due to the development of the ND mechanism (i.e., $Z$ goes to infinity). The horizontal boundary of the ND region is due to the fact that the minimum scale height is regulated by the KH instability, independently of $\alpha_{\rm mid}$ (Eq.~\ref{eq_Hp_KH}). The diagonal boundary is a function of $\alpha_{\rm mid}$ (Eq.~\ref{eq_Hp_tur}). For the analytical derivations of these boundaries, we refer to \citetalias{Hyo21a}.

The parameters controlling the ND instability are as follows. A larger $\alpha_{\rm acc}$ yields a smaller gas surface density in a steady accretion disk (Eq.~(\ref{eq_sigma_g})), thus leading to a higher solid-to-gas ratio for a fixed $\tau_{\rm s}$ and favoring the ND mechanism. Alternatively, for a fixed $\alpha_{\rm acc}$, a smaller $\tau_{\rm s}$  leads to a less efficient radial drift of pebbles, leading to a higher concentration of pebbles in the disk midplane, also favoring the ND mechanism. These arguments help to account qualitatively for the variations of the extent of the ND instability space in Fig.~\ref{fig_parameter}. We note that the analytical predictions from Eq.~(\ref{eq_ND_cri}) (see also Eq.~(20) in \citetalias{Hyo21a}) for $\alpha_{\rm acc}=10^{-2}$ and $\tau_{\rm s}=0.01$ slightly deviate from the direct solutions (see the top left panel of Fig.~\ref{fig_parameter}). This is because the analytical arguments were derived based on the assumption of $\alpha_{\rm acc} \ll \tau_{\rm s}$ (see more details in  \citetalias{Hyo21a}), while here $\alpha_{\rm acc}=\tau_{\rm s}$.

Our numerical simulations including a modeled dead zone with the pebble drift can correspond to Fig.~\ref{fig_parameter} in the following way: First, in a steady accretion disk, the pebble-to-gas mass flux, $F_{\rm p/g}$, is conserved without sublimation and condensation. Then, when drifting pebbles enter a dead zone, $\alpha_{\rm mid}/\alpha_{\rm acc}$ decreases (i.e., Eq~(\ref{eq_alpha_mid})). Thus, a given pebbles flux (i.e., pebbles) moves horizontally from right to left in Fig.~\ref{fig_parameter}. When pebbles cross the diagonal boundary into the red-shaded region in Fig.~\ref{fig_parameter}, the ND mechanism is triggered (Here, $Z=1$ is critical for driving the ND mechanism; see \citetalias{Hyo21a}). This is the reason why the critical condition -- the on/off of the ND mechanism -- does not depend on the radial shape of a dead zone \citepalias{Hyo21a}.

\subsection{Planetesimal belt formed by the ND mechanism} \label{sec_overall}

\begin{figure*}[h]
	\centering
	\resizebox{\hsize}{!}{ \includegraphics{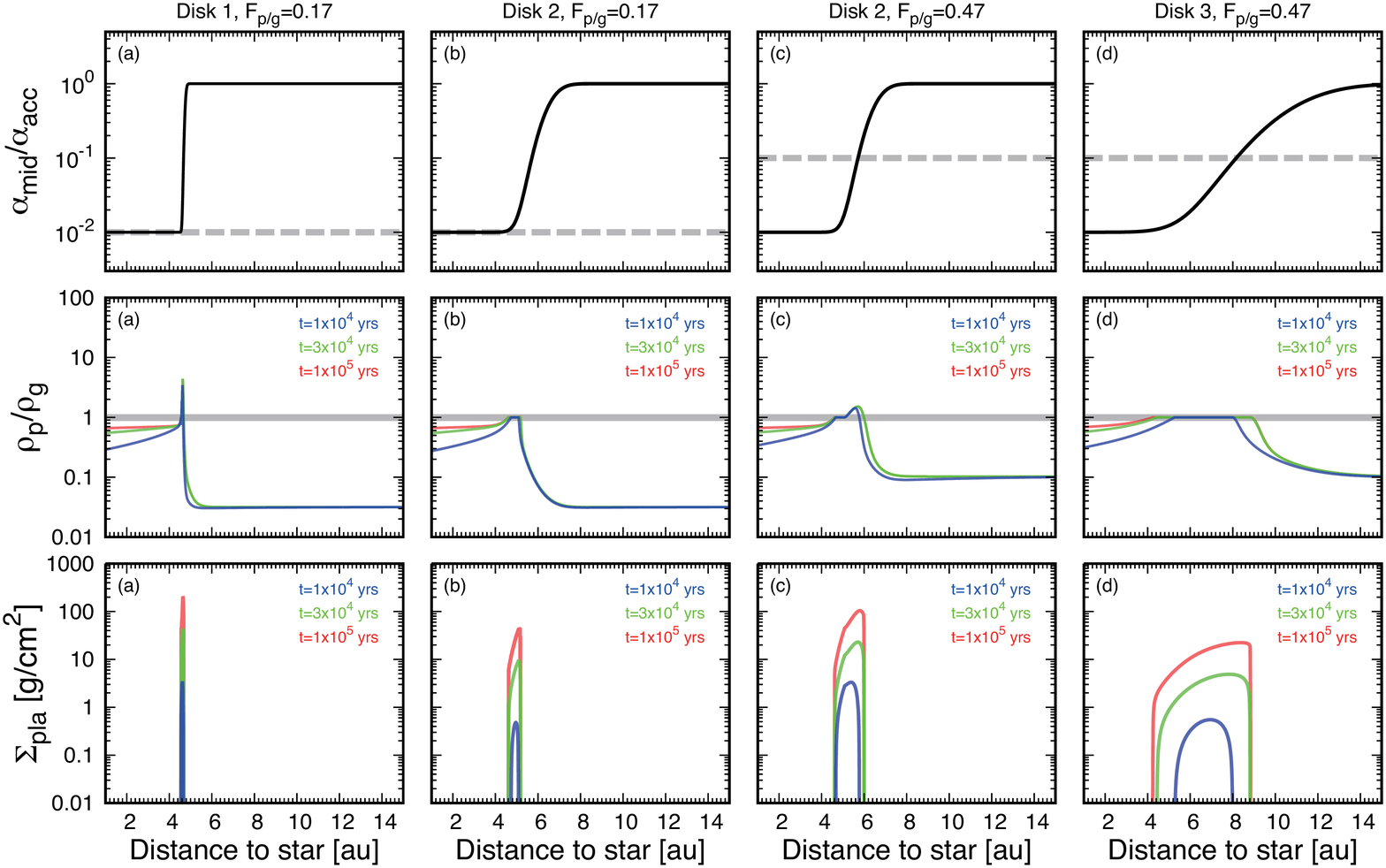} }
	\caption{Overall results of our numerical simulations for different combinations of the disk structure and $F_{\rm p/g}$. Top panels: given disk structures (the black lines; see Table \ref{table_disk}). Middle panels: resultant midplane pebble-to-gas ratio of the spatial density, $\rho_{\rm p}/\rho_{\rm g}$. Bottom panels: resultant planetesimal surface density formed by the ND mode and SI (Eq.~(\ref{eq_planetesimal})). The blue, green, and red lines indicate time evolution. By $t \simeq 1\times 10^{5}$ years, the system reaches a steady-state in $\rho_{\rm p}/\rho_{\rm g}$ and then planetesimals keep forming in a steady width. The dashed gray lines in the top panels represent analytically derived critical $\alpha_{\rm mid}/\alpha_{\rm acc}$ for a given $F_{\rm p/g}$ (Eq.~(\ref{eq_ND_cri})), indicating that disks of smaller $\alpha_{\rm mid}/\alpha_{\rm acc}$ than this critical value develop the ND mode. The gray lines in the middle panels represent the critical $\rho_{\rm p}/\rho_{\rm g}$ above which the ND mode occurs (i.e., $\rho_{\rm p}/\rho_{\rm g}=1$). Here, $\alpha_{\rm acc}=10^{-2}$, $\tau_{\rm s}=0.1$, $\dot{M}_{\rm g} = 10^{-8}\,M_{\odot}$\,yr$^{-1}$, and $k_{\rm SI}=10^{-3}$ are used.}
\label{fig_planetesmial}
\end{figure*}

Figure \ref{fig_planetesmial} shows the disk structures (top panels), the midplane pebble-to-gas ratio (middle panels), and the resultant planetesimal surface density formed by the ND mode (bottom panels) for different combinations of disk structures (Table \ref{table_disk}) and $F_{\rm p/g}$. All cases developed the ND mode and formed planetesimals. The dashed gray lines in the top panels show the analytically derived critical $\alpha_{\rm mid}/\alpha_{\rm acc}$ for a given $F_{\rm p/g}$ (Eq.~\ref{eq_ND_cri}), indicating that disks having smaller $\alpha_{\rm mid}/\alpha_{\rm acc}$ than the critical values are expected to develop the ND mode. The critical $\alpha_{\rm mid}/\alpha_{\rm acc}$ is analytically given as \citepalias[][their Eq. (22)-(23)]{Hyo21a}. 
\begin{align}
	\frac{\alpha_{\rm mid,crit}}{\alpha_{\rm acc}} &\equiv \left( \frac{3 F_{\rm p/g}}{C_{\rm \eta}} \right)^2 \alpha_{\rm acc} \tau_{\rm s}^{-1} \nonumber \\
	& \simeq 4.76 \times 10^{-3} \left( \frac{F_{\rm p/g}}{0.1} \right)^2 \left( \frac{C_{\rm \eta}}{11/8} \right)^{-2} \left( \frac{\alpha_{\rm acc}}{10^{-2}} \right) \left( \frac{\tau_{\rm s}}{0.1} \right)^{-1}.
\label{eq_ND_cri}
\end{align}
This criterion is in very good accordance with the numerical simulations. The gray lines in the middle panel show the critical $\rho_{\rm p}/\rho_{\rm g} = 1$ above which the ND mode is developed. We note that Eq.~(\ref{eq_ND_cri}) is derived under the assumption of $\alpha_{\rm acc} \ll \tau_{\rm s}$. 

The resultant planetesimal belts have diverse widths and profiles depending on the disk structure and the value of $F_{\rm p/g}$ (Fig.~\ref{fig_planetesmial}). The leftmost and rightmost two panels have the same $F_{\rm p/g}$ but have different disk structures, respectively. The middle two panels have the same disk structure but have different $F_{\rm p/g}$.

Early in its evolution when the ND mode has just begun, the pile-up of pebbles and thus planetesimal formation start only at the vicinity of the critical radial distance where the ND occurs for a given $F_{\rm p/g}$ (Eq.~(\ref{eq_ND_cri}), that is, the radial distance where the dashed gray line and the black line intersect in the top panels in Fig.~\ref{fig_planetesmial}). As pile-up continues, diffusion of pebbles enlarges its width and thus the resultant width of the planetesimal belt correspondingly stretches. 

Because $F_{\rm p/g}$ is kept constant at the outer boundary in our simulation, the system eventually reaches a steady-state in $\rho_{\rm p}/\rho_{\rm g}$ (middle panels) and reaches a steady width of planetesimal belt for a given combination of disk structure and $F_{\rm p/g}$ value (time-evolutions are indicated by colors in Fig.~\ref{fig_planetesmial}). After the system reaches a steady-state in $\rho_{\rm p}/\rho_{\rm g}$, planetesimals keep forming within a fixed radial width.

The dependence on the disk structure (compare leftmost and/or rightmost two panels in Fig.~\ref{fig_planetesmial}) is as follow. When the disk has a radially sharper dead zone (i.e., $\alpha_{\rm mid}/\alpha_{\rm acc}$ changes abruptly with radial direction), the resultant planetesimal belt becomes narrower and its surface density becomes more peaked. 

Regarding the $F_{\rm p/g}$ dependence (compare the middle panels of Fig. ~\ref{fig_planetesmial}), a greater $F_{\rm p/g}$ leads to a wider planetesimal belt. The timescale to reach a steady-state solution for the width of planetesimals belt is longer for a greater $F_{\rm p/g}$. The reason for this is discussed in Sect.~\ref{sec_edge}.

The steady-state $\rho_{\rm p}/\rho_{\rm g}$ value also depends on the disk structure and $F_{\rm p/g}$. For the streaming instability to operate, $\rho_{\rm p}/\rho_{\rm g} \geq 1$ is required (Eq.~(\ref{eq_planetesimal})). The timescale to reach a steady-state and the steady-state solution itself are regulated by a complex interplay between the inclusion of new drifting pebbles, the radial diffusion of pebble pile-ups, and the conversion from pebbles to planetesimals via SI (Eq.~(\ref{eq_planetesimal})). Thus, whether steady-state values of $\rho_{\rm p}/\rho_{\rm g}$ become $\geq 1$ depends on these conditions, $k_{\rm SI}$, and $\tau_{\rm s}$ (see Sect.~\ref{sec_kSI} and Sect.~\ref{sec_tau}).

\subsection{Dependence on $k_{\rm SI}$} \label{sec_kSI}

\begin{figure*}[h]
	\centering
	\resizebox{0.6\hsize}{!}{ \includegraphics{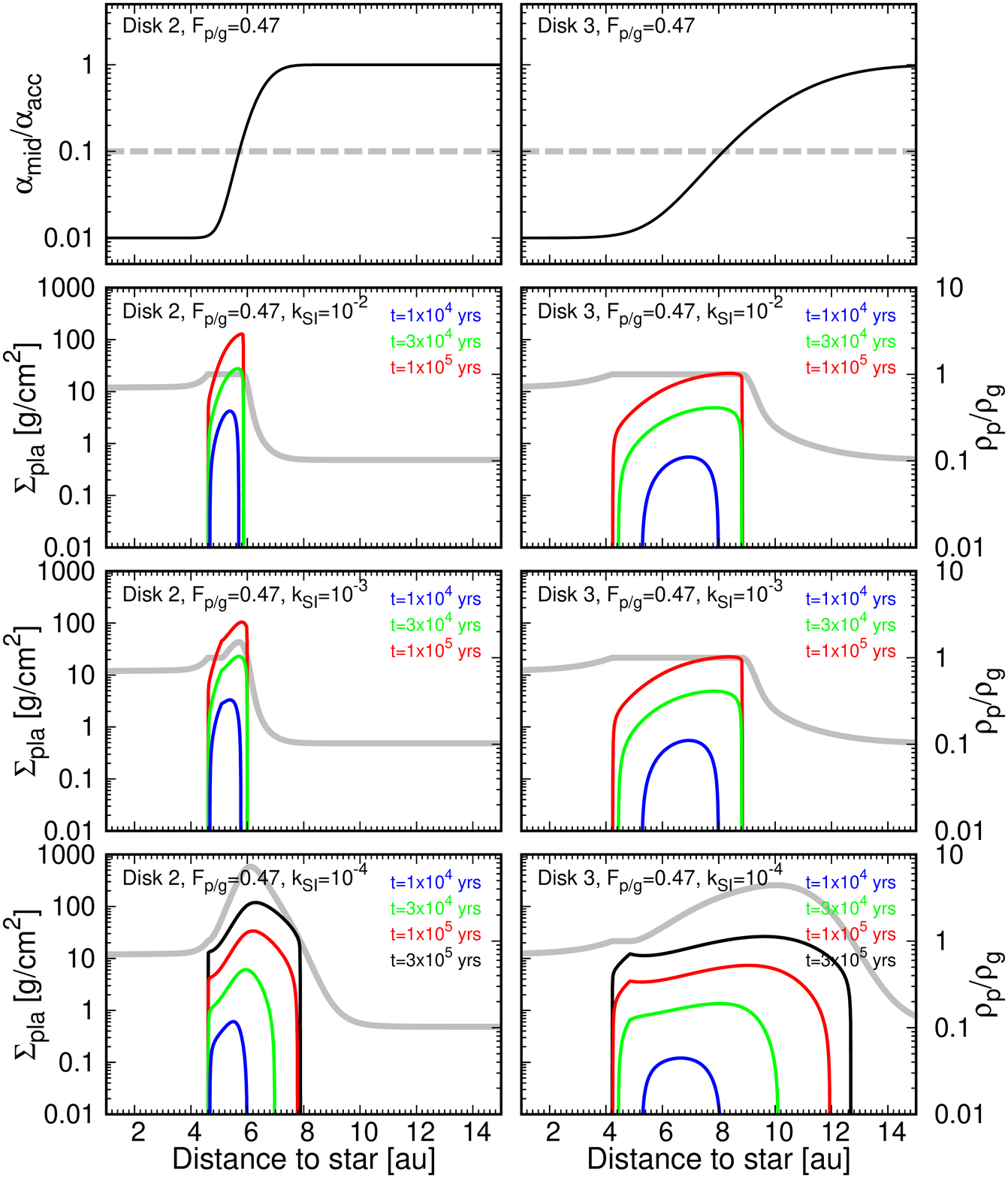} }
	\caption{Dependence of the resultant planetesimal belt on $k_{\rm SI}$ . Left and right panels: cases of different given disk structures (Disk 2 and Disk 3, respectively; Table \ref{table_disk}). Top panels: given disk structures. Bottom three panels: the same combination of $F_{\rm p/g}$ and the disk structure but different $k_{\rm SI}$. The dashed gray lines in the top panels show the analytically derived critical $\alpha_{\rm mid}/\alpha_{\rm acc}$ for a given $F_{\rm p/g}$, indicating that disks having smaller $\alpha_{\rm mid}/\alpha_{\rm acc}$ than the critical one are expected to develop the ND mode (no dependence on $k_{\rm SI}$; Eq.~(\ref{eq_ND_cri})). The blue, green, red, and black lines represent time evolution. By $t \simeq 3 \times 10^{5}$ years, all cases reach a steady-state in $\rho_{\rm p}/\rho_{\rm g}$ and the planetesimals keep forming in a ring with a constant width. The gray lines represent the midplane pebble-to-gas ratio, $\rho_{\rm p}/\rho_{\rm g}$, at the steady-state shown in the right axis. Here, $\tau_{\rm s}=0.1$ and $\dot{M}_{\rm g}=10^{-8}\,M_{\odot}$\,yr$^{-1}$ are used.}
\label{fig_kSI}
\end{figure*}

The efficiency of planetesimal formation via the streaming instability, that is, the ratio of pebbles that are turned to planetesimals within one orbital period, is characterized by the  $k_{\rm SI}$ coefficient as defined in  Eq.~\eqref{eq_planetesimal}. Here, we show the dependence of the resultant planetesimal belt on $k_{\rm SI}$.

Without any planetesimal formation, a runaway pile-up of pebbles (i.e., ND mode) starting at a critical radial distance continuously propagates radially outward (see Figure 2 in \citetalias{Hyo21a}). Including planetesimal formation via SI would change this picture as the final pile-up would be regulated by the relative efficiency of planetesimal formation via SI and that of the pile-up of pebbles via ND mode. 

Figure \ref{fig_kSI} shows the dependence of the resultant planetesimal belt formed via SI on $k_{\rm SI}$. The left and right panels are the cases of Disk 2 and Disk 3, respectively (Table \ref{table_disk}). Disk 2 case has a sharper radial change in $\alpha_{\rm acc}/\alpha_{\rm mid}$ than that of Disk 3 (top panels). Here, $F_{\rm p/g}=0.47$ is used for both cases. The results of the $k_{\rm SI}$ dependence can be categorized into two types. 

First, the results can be converged for a large $k_{\rm SI}$ case depending on the disk structure (see cases of Disk 3 with $k_{\rm SI}=10^{-2}$ and $k_{\rm SI}=10^{-3}$). In this case, the timescale of planetesimal formation is shorter than that of pile-up by the ND mechanism to reach $\rho_{\rm p}/\rho_{\rm g}>1$. Thus, the steady-state in the midplane pebble-to-gas ratio is $\rho_{\rm p}/\rho_{\rm g} \sim 1$, which is the minimum value required for the SI to operate we set in this study (the gray lines in Fig.~\ref{fig_kSI}). Thus, for high-enough $k_{\rm SI}$ values, the results are independent of $k_{\rm SI}$.

Second, the widths of planetesimal belt and the timescales to reach a steady-state in $\rho_{\rm p}/\rho_{\rm g}$ become wider and longer for a smaller $k_{\rm SI}$ (see cases Disk 2 with $k_{\rm SI}=10^{-2}, 10^{-3}$, and $10^{-4}$, and cases of Disk 3 with $k_{\rm SI}=10^{-3}$ and $10^{-4}$). In this case, the timescale of planetesimal formation is longer than that of pile-up to reach $\rho_{\rm p}/\rho_{\rm g}=1$ depending on the $k_{\rm SI}$ value. Thus, for smaller $k_{\rm SI}$ values, the steady-state value of $\rho_{\rm p}/\rho_{\rm g}$ becomes larger than unity (i.e., $\rho_{\rm p}/\rho_{\rm g} > 1$; the gray lines in Fig.~\ref{fig_kSI}).

\subsection{Dependence on $\dot{M}_{\rm g}$} \label{sec_dotMg}

\begin{figure*}[h]
	\centering
	\resizebox{\hsize}{!}{ \includegraphics{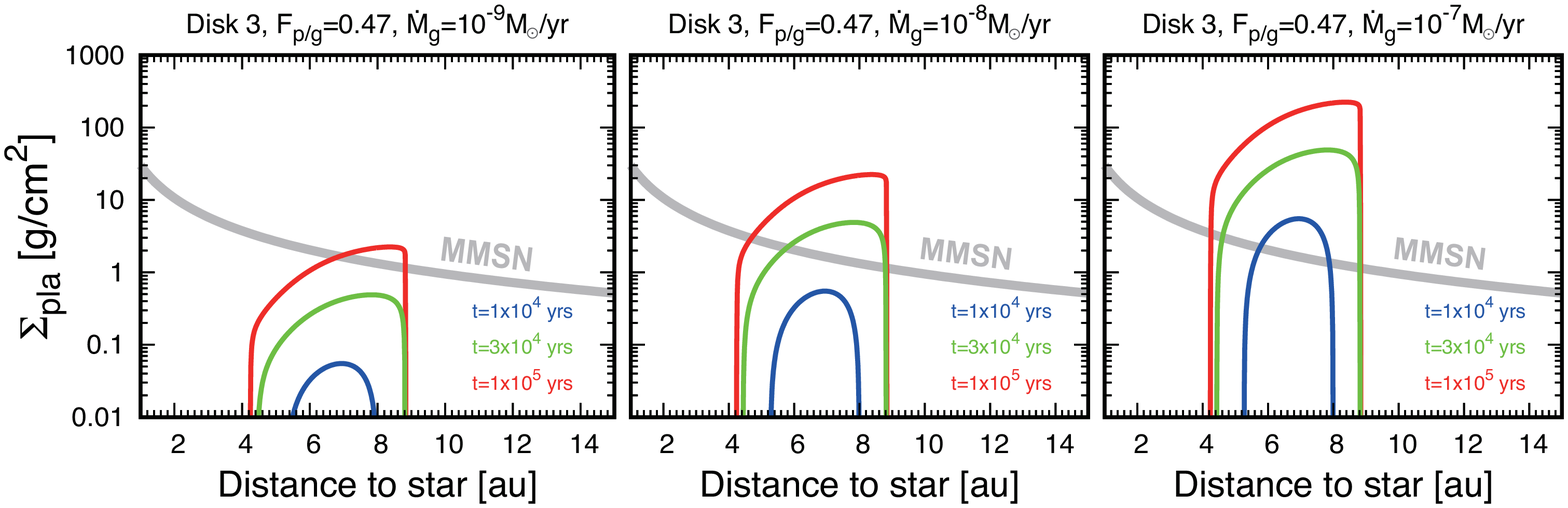} }
	\caption{Dependence of the resultant planetesimal belt on $\dot{M}_{\rm g}$. Left to right panels are the cases of $\dot{M}_{\rm g}=10^{-9}, 10^{-8}$, and $10^{-7}\,M_{\odot}$\,yr$^{-1}$, respectively. The gray lines represent the surface density of solids in the classical minimum mass solar nebula model (MMSN; $\Sigma_{\rm pla}=30(r/1{\, \rm au})^{-3/2}$\,g\,cm$^{-2}$) of \cite{Wei77} for reference. By $t \simeq 1\times 10^{5}$ years, the system reaches a steady-state in $\rho_{\rm p}/\rho_{\rm g}$ and width of the planetesimal ring. Here, $\alpha_{\rm acc}=10^{-2}$, $\tau_{\rm s}=0.1$, and $k_{\rm SI}=10^{-3}$ are used.}
\label{fig_Mdot}
\end{figure*}

In this subsection, we study the dependence on $\dot{M}_{\rm g}$ in the resultant planetesimal belt. Using the classical $\alpha$-disk model and $F_{\rm p/g}$, the pebble surface density is characterized by $\Sigma_{\rm p} \propto F_{\rm p/g} \times \dot{M}_{\rm g}$. Thus, a higher value of $\dot{M}_{\rm g}$ also yields a larger absolute amount of pebbles in the disk.

The criteria and conditions of the ND mechanism are described in nondimensional forms by using $F_{\rm p/g} \equiv \dot{M}_{\rm p}/\dot{M}_{\rm g}$ (see Eq.~\ref{eq_ND_cri}). Thus, the resultant absolute planetesimal surface density linearly scales with $\dot{M}_{\rm g}$ as $\Sigma_{\rm pla} \propto \Sigma_{\rm p} \propto \dot{M}_{\rm p} \propto F_{\rm p/g} \times \dot{M}_{\rm g}$.

Figure \ref{fig_Mdot} shows the resultant planetesimal belts for cases with the same $F_{\rm p/g}$ value and disk structure but for different $\dot{M}_{\rm g}$ values. As discussed above, a larger amount of planetesimals is formed for a larger $\dot{M}_{\rm g}$, and it scales with $\dot{M}_{\rm g}$; the profile and width are the same but only the absolute value scales with $\dot{M}_{\rm g}$.

\subsection{Edges of the planetesimal belt} \label{sec_edge}
In the previous subsections, we discussed the dependences of the widths and profiles of the planetesimal belts formed by the ND mechanism on $k_{\rm SI}$ and $\dot{M}_{\rm g}$. We numerically demonstrated that the resultant planetesimal distribution finally reaches a steady-state to have a fixed width for a constant $F_{\rm p/g}$.

One may notice that the inner edge of the planetesimal belt is nearly independent on $k_{\rm SI}$ and $\dot{M}_{\rm g}$ for a given disk structure (Figs.~\ref{fig_kSI} and \ref{fig_Mdot}). The inner edge is characterized by the disk structure; it is located at the vicinity of the outermost radial distance where $\alpha_{\rm mid}/\alpha_{\rm acc}$ no longer decreases with decreasing $r$. 

The ND pile-up starts at a greater radial distance, defined by $r_{\rm ND,start}$ (i.e., characterized by Eq.~(\ref{eq_ND_cri}) where a critical $\alpha_{\rm mid}/\alpha_{\rm acc}$ is a function of $F_{\rm p/g,given}$ we set at the outer boundary). Because accumulating pebbles radially diffuse, a fraction of pebbles diffuse inward and the pebble mass flux at $r < r_{\rm ND,start}$ smoothly propagates to have $F_{\rm p/g} < F_{\rm p/g,given}$. At $r < r_{\rm ND,start}$, the required $F_{\rm p/g}$ for the ND mode to develop is smaller as $\alpha_{\rm mid}/\alpha_{\rm acc}$ decreases with decreasing $r$ and thus the ND mode smoothly keeps being developed at $r < r_{\rm ND,start}$ until $\alpha_{\rm mid}/\alpha_{\rm acc}$ reaches its minimum value. 

The outer edge of the planetesimal belt depends on $k_{\rm SI}$ and the disk structure (Figs~\ref{fig_planetesmial} and \ref{fig_kSI}). Smoother radial change in $\alpha_{\rm mid}/\alpha_{\rm acc}$ and/or smaller $k_{\rm SI}$ leads to the outer edges of the planetesimal belt more radially distant for a given $F_{\rm p/g}$ as accumulating pebbles more efficiently diffuse outward while turning into planetesimals. The radial change in $\alpha_{\rm mid}$ results in an additional diffusion flux to the case of a constant diffusivity (see Eq.~(\ref{eq_sigma_peb}) which includes the term of the dependence on $(\partial D_{\rm p}(\alpha_{\rm mid}) / \partial r) \times (\partial Z_{\rm \Sigma}/\partial r)$ where $Z_{\rm \Sigma} \equiv \Sigma_{\rm p}/\Sigma_{\rm g}$). A smaller value of $k_{\rm SI}$ indicates a less effective conversion from accumulating pebbles to planetesimals, and thus the pebble pile-up propagates outward.

\subsection{Cumulative planetesimal mass formed by the ND mode} \label{sec_mass}

\begin{figure*}[h]
	\centering
	\resizebox{\hsize}{!}{ \includegraphics{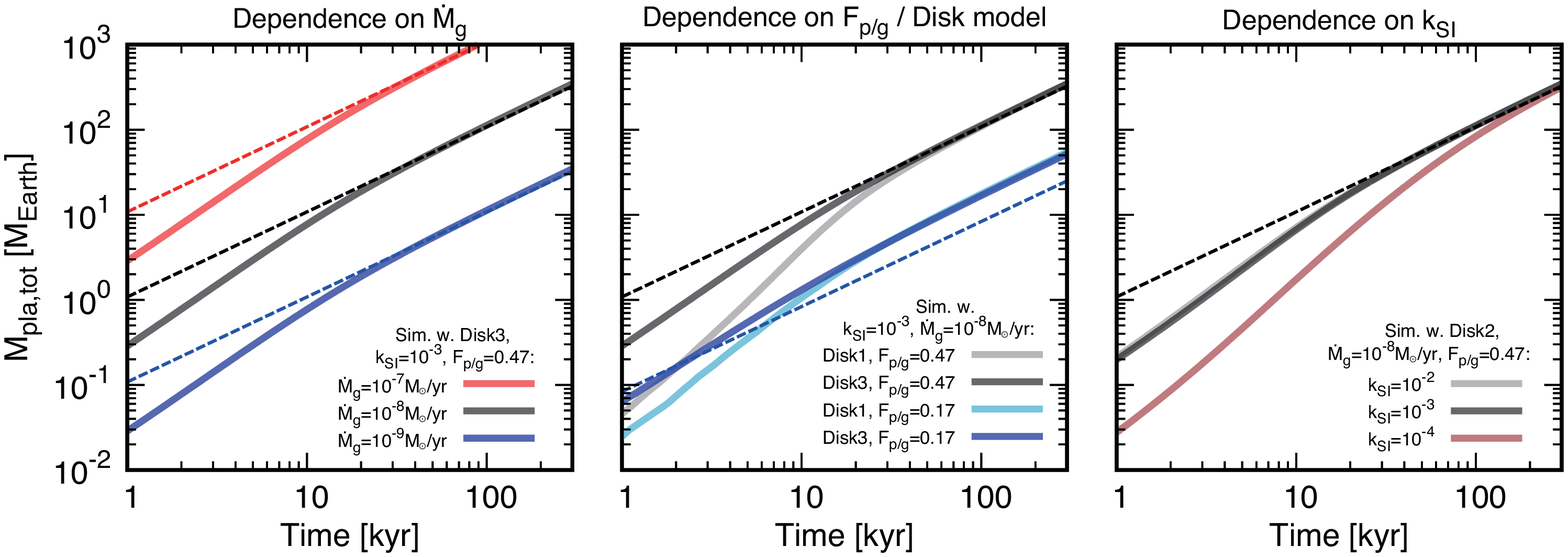} }
	\caption{Cumulative mass of planetesimals, $M_{\rm pla,tot}$, formed by the ND as a function of time. The thin dashed lines are analytically derived cumulative masses of planetesimals (Eq.~(\ref{eq_pla_ana})). The thick solid lines are the results of numerical simulations. Here, we sorted numerical data so that planetesimal formation starts at $t=0$\,kyr (i.e., the time at which $\rho_{\rm d}/\rho_{\rm g}$ exceeds 1, which should depend on the parameters in the numerical simulations, is always shifted to $t = 0$\,kyr).  Left panel: dependence on $\dot{M}_{\rm g}$ where $\dot{M}_{\rm g}(=10^{-7}, 10^{-8}$, and $10^{-9}\,M_{\odot}$\,yr$^{-1}$) are a variable while the disk structure (Disk 3) and $F_{\rm p/g}=0.47$ are kept the same. Middle panel: dependence on disk structure and $F_{\rm p/g}$ where disk structure (Disk 1 or Disk 3; see table \ref{table_disk}) and $F_{\rm p/g}(=0.17$ or $0.47$) are variables while keeping the same $\dot{M}_{\rm g}=10^{-8} M_{\odot}$ yr$^{-1}$. Right panel: dependence on $k_{\rm SI}$ where $k_{\rm SI}=10^{-2}, 10^{-3},$ and $10^{-4}$ while having the same disk structure (Disk 2), $F_{\rm p/g}=0.47$, and  $\dot{M}_{\rm g}=10^{-8}\,M_{\odot}$\,yr$^{-1}$. After around $\sim10-100$\,kyr, depending on the parameters, the rate of planetesimal formation can be described by the analytical arguments (dashed lines); that is, the system reaches a steady width of the planetesimal belt (Sect.~\ref{sec_mass}). In this study, $F_{\rm p/g}$ is kept constant (i.e., infinite solid mass budget). We note, however, that the maximum total mass of planetesimals in a real system should be regulated by the mass budget of solids (see Sect.~\ref{sec_mass}). Here, $\alpha_{\rm acc}=10^{-2}$ and $\tau_{\rm s}=0.1$ are used.}
\label{fig_rate}
\end{figure*}

Here, we aim to understand the total mass of planetesimals contained within the belt. Below, we first analytically derive the rate of planetesimal formation when the system reaches a steady-state in $\rho_{\rm p}/\rho_{\rm g}$. Then, the validity of our analytical arguments is checked by comparing them with the numerical simulations (Fig.~\ref{fig_rate}).

Following the discussion in Sect.~\ref{sec_edge}, a reduced pebble mass flux propagates continuously inward from the radial point where the ND mechanism initially takes place at $r = r_{\rm ND,start}$ for a given $F_{\rm p/g,given}$. The development of the ND mode correspondingly propagates inward because $\alpha_{\rm mid}/\alpha_{\rm acc}$ decreases with decreasing $r$. The ND mode develops until it reaches the minimum $\alpha_{\rm mid}/\alpha_{\rm acc}$ point (i.e., the minimum midplane turbulence of $\alpha_{\rm mid,min}$). At this radial distance, $F_{\rm p/g}$ becomes its minimum for the ND mode to develop. This minimum value, $F_{\rm p/g,ND,min}$, is given from Eq.~(\ref{eq_ND_cri}) as
\begin{align}
	F_{\rm p/g,ND,min} & = \frac{\left( \alpha_{\rm mid,min}\tau_{\rm s}\right)^{1/2} C_{\rm \eta}}{3\alpha_{\rm acc}} \nonumber \\
	& \simeq 0.145 \left( \frac{\alpha_{\rm acc}}{10^{-2}} \right)^{-1} \left( \frac{\alpha_{\rm mid,min}}{10^{-4}} \right)^{1/2} \left( \frac{\tau_{\rm s}}{0.1} \right)^{1/2} \left( \frac{C_{\rm \eta}}{11/8} \right).
	\label{eq_Fpg_min}
\end{align}

Therefore, the pebble mass flux is reduced from $F_{\rm p/g}=F_{\rm p/g,given}$ at the outer boundary to $F_{\rm p/g,ND,min}$. This difference leads to the rate of pebble pile-up and is eventually converted to planetesimals via SI. To estimate the rate of planetesimal formation via the ND mechanism, the following mass balance is considered. From the outer region of the disk, pebbles with $F_{\rm p/g,given}$ are supplied to the ND region within a dead zone, and a fraction of the supplied pebbles are flowed out further inward from the ND region with $F_{\rm p/g,ND,min}$. Thus, the cumulative mass of planetesimals formed by the ND mode during the time interval of $\Delta t$ is given as
\begin{align}
	M_{\rm pla,tot} &= F_{\rm p/g}^{\rm in/out} \dot{M}_{\rm g} \times \Delta t \nonumber \\
	& = \left( F_{\rm p/g,given} - F_{\rm p/g,ND,min} \right) \dot{M}_{\rm g} \times \Delta t  \nonumber \\
	& \simeq 3.3 \,{\Mearth} \left( F_{\rm p/g} - F_{\rm p/g,ND,min} \right) \left( \frac{\dot{M}_{\rm g}}{10^{-8} \, M_{\rm \odot} \, {\rm yr}^{-1}} \right) \left( \frac{\Delta t} {1 \, {\rm kyr}} \right) ,
\label{eq_pla_ana}
\end{align}
where $F_{\rm p/g}^{\rm in/out} \equiv F_{\rm p/g,given} - F_{\rm p/g,ND,min}$ is the net mass flux supplied to the ND region. 

Importantly, this analytical estimation indicates that the total mass of planetesimals formed via the ND mechanism through SI is independent on the disk's radial structure and $k_{\rm SI}$, that is, irrespective of the smooth or sharp radial change in $\alpha_{\rm mid}/\alpha_{\rm acc}$. Instead, it is characterized by $F_{\rm p/g,given}$, the minimum value of $\alpha_{\rm mid}/\alpha_{\rm acc}$ through $F_{\rm p/g,ND,min} $ (Eq.~(\ref{eq_Fpg_min})), and $\dot{M}_{\rm g}$.

The maximum total mass of planetesimals should be regulated by the mass budget of solid in a disk. Considering a disk having a disk-to-star mass ratio of $f_{\rm disk/star}$ and a disk's solid-to-gas mass ratio of $f_{\rm solid/gas}$, the maximum solid mass is $M_{\rm solid} = f_{\rm disk/star} f_{\rm solid/gas} \,{\Msun} \simeq 3 \times 10^{5} f_{\rm disk/star} f_{\rm solid/gas} \,{\Mearth}$ for solar mass star. Using the efficiency of the planetesimal formation in Eq~(\ref{eq_pla_ana}), the maximum total planetesimal mass, $M_{\rm pla,max} $, is given as 
\begin{align}
	& M_{\rm pla,max} = \left( \frac{F_{\rm p/g}^{\rm in/out}}{F_{\rm p/g,given}} \right) M_{\rm solid} \nonumber \\
	&\simeq 30  \left( \frac{F_{\rm p/g}^{\rm in/out}}{F_{\rm p/g,given}} \right) \left( \frac{f_{\rm disk/star}}{0.01} \right)  \left( \frac{f_{\rm solid/gas}}{0.01} \right) \,{\Mearth}\nonumber \\	
	&= 30 \left( \frac{F_{\rm p/g,given} - F_{\rm p/g,ND,min}}{F_{\rm p/g,given}} \right)  \left( \frac{f_{\rm disk/star}}{0.01} \right)  \left( \frac{f_{\rm solid/gas}}{0.01} \right) \,{\Mearth} .
\label{eq_pla_max}
\end{align}
Accordingly, for the classical minimum mass solar nebula (MMSN) model of $f_{\rm disk/star} \sim 0.01$ \citep{Wei77} and for the interstellar medium value of $f_{\rm solid/gas} \sim 0.01$ \citep{Boh78}, $M_{\rm pla,max} \sim 20 \,{\Mearth}$ and $\sim 4.5\,{\Mearth}$ for $F_{\rm p/g,given} = 0.47$ and $0.17$ with $\alpha_{\rm acc}=10^{-2}$ and $\alpha_{\rm mid}=10^{-4}$, respectively. We note that the MMSN (i.e., $M_{\rm solid} \simeq 30\Mearth$) is just a single benchmark in the limit of the smallest mass budget. The ND mechanism works better for a larger mass budget. Recent studies in the context of drifting pebbles \citep[e.g.,][]{Lam14,Ida16b,Bit19,Dra21} considered the cases of a more abundant solid mass budget (e.g., $M_{\rm solid} \simeq 650 \Mearth$ in \cite{Dra21}), where more planetesimals are expected to be formed by the ND mechanism.

Figure \ref{fig_rate} shows the cumulative mass of planetesimals, $M_{\rm pla,tot}$, formed by the ND as a function of time. Here, $t=0$ is set to be the point where $M_{\rm pla,tot}=0$ in the numerical simulations, that is, the time at which $\rho_{\rm d}/\rho_{\rm g}$ exceeds 1, which should depend on the parameters in the numerical simulations, is always shifted to $t = 0$. The thin dashed lines are the analytical estimates (Eq.~(\ref{eq_pla_ana})) and the thick solid lines are numerical results. The rate of planetesimal formation seen in the numerical simulations can be divided into two phases -- the early nonlinear phase and the successive linear phase.

In the early stage of the evolution (during the first $\sim 10-100$\,kyr after the start of planetesimal formation for $k_{\rm SI} \gtrsim 10^{-3}$), deviations are seen between analytical estimates and numerical results with an early evolution that is nonlinear. During this early phase, the width of the planetesimal belt changes as pebbles affected by the ND mechanism diffuse radially inward and outward, thus affecting the pebble surface density profile (e.g., Fig.~\ref{fig_planetesmial}). A fraction of incoming pebbles are used to increase the pebble surface density in the ND region toward a steady-state, while only a fraction of pebbles are used to form planetesimals. The efficiency of the planetesimal formation can then be significantly smaller than that predicted in steady-state by Eq~(\ref{eq_pla_ana}). Accretion peaks (with high $F_{\rm p/g}$ values) that are too short ($\Delta t \ll 10-100$\,kyr) hence would not yield an efficient formation of planetesimals.

After about $10-100$\,kyr since the start of planetesimal formation, the system reaches a steady-state and the efficiency of the planetesimal formation is linear and can be described by Eq~(\ref{eq_pla_ana}). A small mismatch is still seen even after the system reaches the steady-state for the $F_{\rm p/g}=0.17$ cases (middle panel). This is probably because this value is close to $F_{\rm p/g,ND,min} \simeq 0.15$ so that even a small error in the analytical estimation of $F_{\rm p/g,ND,min}$ (e.g., neglecting the effects of diffusion) may lead to a noticeable difference. We confirmed that this deviation becomes smaller for greater $F_{\rm p/g}$ values.

A non-negligible total mass of planetesimals (more than Earth mass) could be formed for a disk having $F_{\rm p/g} \gtrsim 0.1$ for a duration $\gtrsim 10-100$\,kyr and a gas mass flux of $\gtrsim 10^{-8}\,{\Msun}$\,yr$^{-1}$, and $\alpha_{\rm mid}/\alpha_{\rm acc} \simeq 10^{-2}$ with $\alpha_{\rm acc}=10^{-2}$ at $r \lesssim 10$\,au. Dependencies on the other factors are discussed below.

At the steady-state, as expected from the analytical estimation, the total mass of planetesimals formed within the planetesimal belt does not depend on the width and profile of the belt (i.e., irrespective of the dead zone structure and of $k_{\rm SI}$). Conversely, in the early nonlinear phase of the evolution before reaching the steady-state in $\rho_{\rm p}/\rho_{\rm g}$ (when the width of the planetesimal belt is still changing; see Figs.~\ref{fig_planetesmial}-\ref{fig_Mdot}), the rate of planetesimal formation depends on the disk structures and $k_{\rm SI}$ (see also discussion of the outer edge of the planetesimal belt in Sect.~\ref{sec_edge}). 

The dependencies on $\dot{M}_{\rm g}$, $F_{\rm p/g}$, and $k_{\rm SI}$ are as follows. (1) The total planetesimal mass is linearly scaled with $\dot{M}_{\rm g}$ (left panel in Fig.~\ref{fig_rate}; see also Sect.~\ref{sec_dotMg}). (2) A greater $F_{\rm p/g}$ results in a larger amount of total planetesimal mass (middle panel in Fig.~\ref{fig_rate}; see also Sect.~\ref{sec_edge}). (3) For a smaller SI efficiency (i.e., smaller $k_{\rm SI}$), the timescale to reach the steady-state in $\rho_{\rm p}/\rho_{\rm g}$ with a fixed width of planetesimal belt is longer (right panel in Fig.~\ref{fig_rate}; see also Sect.~\ref{sec_kSI} and Fig.~\ref{fig_kSI}).

\subsection{Dependence on $\tau_{\rm s}$} \label{sec_tau}

\begin{figure*}[h]
	\centering
	\resizebox{\hsize}{!}{ \includegraphics{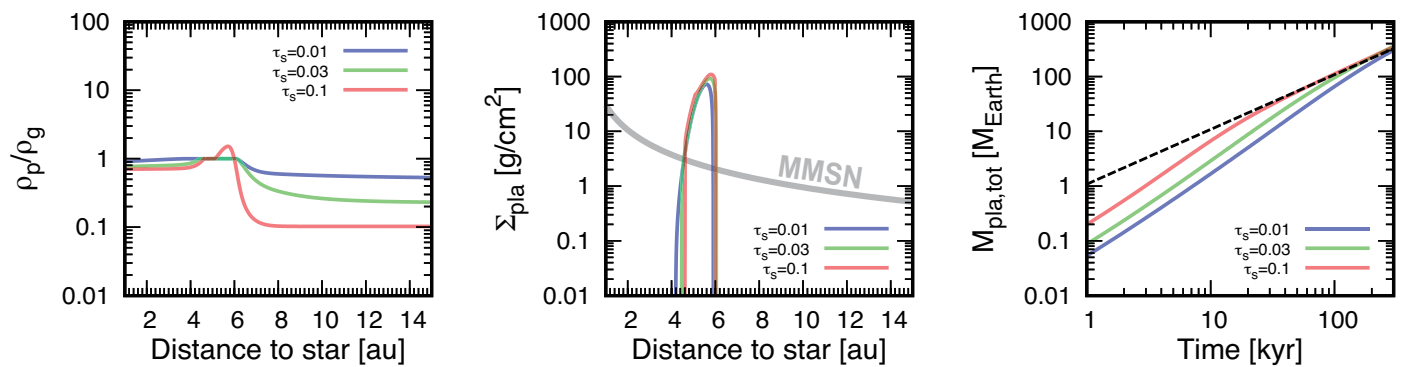} }
	\caption{Dependence of the resultant planetesimal belt on the Stokes number, $\tau_{\rm s}$ (blue for $\tau_{\rm s}=0.01$, green for $\tau_{\rm s}=0.03$, and red for $\tau_{\rm s}=0.1$, respectively). Left panel: resultant midplane pebble-to-gas ratio of the spatial density, $Z \equiv \rho_{\rm p}/\rho_{\rm g}$, after the establishment of the steady-state. Middle panel: resultant planetesimal surface density at $t=100$\,kyr since the start of planetesimal formation. The gray line is the classical minimum mass solar nebula model (MMSN; $\Sigma_{\rm pla}=30(r/1{\, \rm au})^{-3/2}$\,g\,cm$^{-2}$) of \cite{Wei77}. Right panel: cumulative mass of planetesimals, $M_{\rm pla,tot}$, formed by the ND mechanism since the start of planetesimal formation. The thin dashed line is analytically derived cumulative mass of planetesimals with $\tau_{\rm s}=0.1$. Here, Disk 2, $\alpha_{\rm acc}=10^{-2}$, $\dot{M}_{\rm g}=10^{-8}\,M_{\odot}$\,yr$^{-1}$, $F_{\rm p/g}=0.47$, and $k_{\rm SI}=10^{-3}$ are used.}
\label{fig_tau}
\end{figure*}

So far, we have considered that pebbles were characterized by a Stokes parameter of $\tau_{\rm s}=0.1$. Here, we additionally considered $\tau_{\rm s}=0.01$ and $0.03$.

From analytical considerations, the critical $\alpha_{\rm mid}$ below which the ND runaway pile-up occurs depends on $\tau_{\rm s}^{-1}$ (Eq.~(\ref{eq_ND_cri})). Alternatively, the critical $F_{\rm p/g}$ above which the ND mode takes place depends on $\tau_{\rm s}^{1/2}$ (Eq.~(\ref{eq_ND_cri})). Thus, a larger $\alpha_{\rm mid}$ (or a smaller $F_{\rm p/g}$) satisfies the ND criterion for a smaller $\tau_{\rm s}$. This is because, for $\alpha_{\rm acc} \lesssim \tau_{\rm s}$, a smaller $\tau_{\rm s}$ increases the pebble surface density ($\Sigma_{\rm p} \propto v_{\rm p}^{-1}$ and Eq.~(\ref{eq_vp})), while its scale height is independently regulated by $\alpha_{\rm mid}$. Its dependence on $\tau_{\rm s}$ is weaker (Eq.~(\ref{eq_Hp_tur})).

Figure~\ref{fig_tau} shows the numerical results and dependence of the resultant planetesimal belt on the Stokes number ($\tau_{\rm s}=0.01, 0.03$, and 0.1 with Disk-2, $\dot{M}_{\rm g}=10^{-8}\,M_{\odot}$\,yr$^{-1}$, $F_{\rm p/g}=0.47$, $\alpha_{\rm acc}=10^{-2}$, and $k_{\rm SI}=10^{-3}$). Although a smaller $\tau_{\rm s}$ more easily satisfies the ND criterion (i.e., with a smaller $F_{\rm p/g}$), the early-phase efficiency of planetesimal formation before the steady-state decreases and the timescale of the pebble pile-up becomes longer ($\sim 10$\,kyr for $\tau_{\rm s}=0.1$ and $\sim 100$\,kyr for $\tau_{\rm s}=0.01$). This is because the efficiency of the pebble pile-up is regulated by the balance between the continuous accumulation of pebbles by absorbing inwardly drifting pebbles, the radial diffusion of the piled-up pebbles, and the conversion from pebbles to planetesimals via SI. For a smaller $\tau_{\rm s}$, the inward drift of pebbles becomes less efficient (Eq.~(\ref{eq_vp})).

The similarity seen in the resultant radial extent of the planetesimal belt can be understood from the leftmost panel of Fig.~\ref{fig_tau}: it shows that the steady-state $\rho_{\rm p}/\rho_{\rm g}$ is constant and $\simeq 1$ independently of $\tau_{\rm s}$ for small $\tau_{\rm s}$ values. Such a similarity is also seen in Fig.~\ref{fig_kSI} for large $k_{\rm SI}$ values (see the middle panels in Fig.~\ref{fig_kSI}). This is because a smaller $\tau_{\rm s}$ requires a longer timescale to pile up as the drift velocity decreases. Its timescale becomes longer than that of the planetesimal formation via SI, settling to $\rho_{\rm p}/\rho_{\rm g} \simeq 1$, which is the 
 value required for the SI to operate. Finally, we note that $F_{\rm p/g,ND,min} \propto \tau_{\rm s}^{1/2}$ (Eq.~(\ref{eq_Fpg_min})) and thus the total mass of planetesimals weakly depends on $\tau_{\rm s}$ (Eq.~(\ref{eq_pla_ana})).

\subsection{Dependence on $\alpha_{\rm acc}$} \label{sec_acc}

\begin{figure*}[h]
	\centering
	\resizebox{\hsize}{!}{ \includegraphics{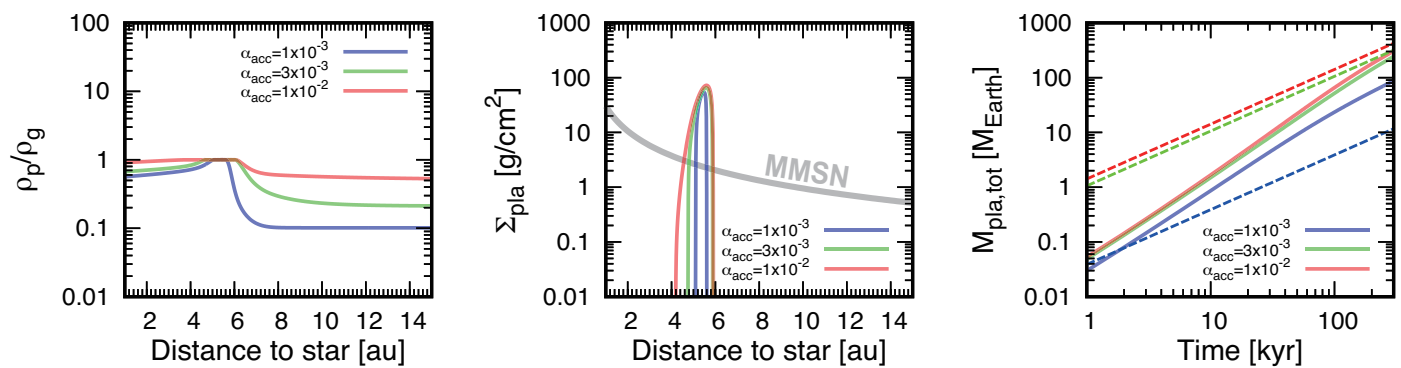} }
	\caption{Same as Fig.~\ref{fig_tau}, but for different $\alpha_{\rm acc}$ (blue for $\alpha_{\rm acc}=1 \times 10^{-3}$, green for $\alpha_{\rm acc}=3 \times 10^{-3}$, and red for $\alpha_{\rm acc}=1 \times 10^{-2}$, respectively). Here, Disk 2, $\tau_{\rm s}=0.01$, $\dot{M}_{\rm g}=10^{-8}M_{\odot}$ yr$^{-1}$, $F_{\rm p/g}=0.47$, and $k_{\rm SI}=10^{-3}$ are used. A large deviation between the analytical estimation (blue dashed line) and the simulation (blue solid line) for $\alpha_{\rm acc}=1 \times 10^{-3}$ is seen in the right panel (see texts for the potential reason).}
\label{fig_acc}
\end{figure*}

In this subsection, we present the dependence on $\alpha_{\rm acc}$. Here, $\alpha_{\rm acc}=1 \times 10^{-3}, 3 \times 10^{-3}$, and $1 \times 10^{-2}$ were used with $\tau_{\rm s}=0.01$. Figure \ref{fig_acc} shows the dependence of the resultant planetesimal belt on $\alpha_{\rm acc}$. Because we used $\tau_{\rm s}=0.01$, the timescale to reach a steady-state becomes longer than with $\tau_{\rm s}=0.1$ ($\sim 100$\,kyr; see the right panel in Fig.~\ref{fig_acc}; see also Sect.~\ref{sec_tau}).

A larger $\alpha_{\rm acc}$ yields a smaller gas surface density (Eq.~(\ref{eq_sigma_g})). This leads to a larger ratio of pebbles to gas midplane density. Thus, the critical $\alpha_{\rm mid}$ for the ND to be triggered becomes larger (Eq.~(\ref{eq_ND_cri})), that is, the ND starts at a larger radial distance for the same dead zone structure. Also, as $Z$ becomes globally higher, the width of the planetesimal formation becomes larger (see the left and middle panels in Fig.~\ref{fig_acc}). The rate of planetesimal formation during the early (nonlinear) phase also depends on $\alpha_{\rm acc}$ (the panel in Fig.~\ref{fig_acc}). This would be because the drift velocity of pebbles is smaller for a smaller $\alpha_{\rm acc}$ and thus the absorption of drifting pebbles into the pebble pile-up is less efficient, delaying the entrance into the steady-state regime.

For the cases of $\alpha_{\rm acc}= 10^{-2}$ and $3 \times 10^{-3}$, the rate of planetesimal formation in steady-state are relatively well described by the analytical relations (see the red and green lines in the right panel of Fig.~\ref{fig_acc} and Eq.~(\ref{eq_pla_ana})). A small deviation is probably related to the fact that Eq.~(\ref{eq_pla_ana}) assumes $\alpha_{\rm acc} \ll \tau_{\rm s}$ while here $\alpha_{\rm acc} \sim \tau_{\rm s}$. For $\alpha_{\rm acc}= 10^{-3}$, a large deviation between the analytical estimation (the blue dashed line) and the numerical simulation (the blue solid line) is seen. This would be because $F_{\rm p/g,ND,min} \simeq 0.46$ is very close to $F_{\rm p/g,given}=0.47$ for $\alpha_{\rm acc}=1 \times 10^{-3}$ with $\tau_{\rm s}=0.01$ -- the same as cases of $F_{\rm p/g,given}=0.17$ seen in the middle panel of Fig.~\ref{fig_rate} -- and the analytical estimation of $F_{\rm p/g,ND,min}$ does not fully include the effects of backreaction and diffusion. This potentially leads to a noticeable difference even from a small error in $F_{\rm p/g,ND,min}$ because $F_{\rm p/g,given} - F_{\rm p/g,ND,mid}$ is used to calculate $M_{\rm pla,tot}$ (see Eq.~(\ref{eq_pla_ana})).

\section{Discussion} \label{sec_discussion}

\subsection{Comparison to other pile-up mechanisms} \label{sec_evolving_disk}
Depending on the disk structures as well as the physicochemical properties of the solids, different mechanisms may be responsible for trapping drifting particles. Here, we compare the ND mechanism with other mechanisms that can potentially pile up solids in disks.

Pressure bumps, radial maxima in the disk gas pressure, naturally lead to convergent motions of solid particles and their pile-up \citep{Whi72,Kre07,Pin12,Cha14,Des18,Dul18}. These pressure bumps may be preferentially formed at the snow line \citep[e.g.,][]{Mul21,Cha19,Cha21} or at the outer edge of a dead zone due to the change of the $\alpha$-viscosity \citep{Pin16}. However, it is still questionable how the pressure bump is efficiently preserved against the backreaction of the solid pile-up \citep[e.g.,][]{Tak16,Kan18}.
 
The pressure bump found by \citet{Pin16} at the edge of a dead zone is caused by the evolution of a standard $\alpha$-disk. However, the evolution of non-standard disk models such as when including disk winds and accounting for vertical variations of angular momentum transport, do not necessarily lead to the formation of such a pressure bump, even in the presence of a dead zone. Also, the feedback, backreaction, from the accumulating pebbles to the gas may smooth out a bump structure. Our study shows that even without a pressure bump, particle pile-up may occur (for appropriate change in the vertical turbulence and pebble-to-gas mass flux) through the ND mechanism. Although this should be studied through dedicated simulations including disk evolution and backreaction effects, in the presence of a pressure bump, the ND mechanism should combine to convergent motions of particles to lead to a more pronounced pile-up of pebbles.

Other mechanisms leading to a pile-up of particles require efficient particle growth and/or fragmentation combined to the backreaction. If these lead to a strong enough modification of the gas profile, dust particles may pile up \citep[e.g.,][]{Gon17,Gar20}. The inner region of disks may also be a favorable location for solids to pile up via a traffic-jam effect, if the fragmentation is efficient enough at a small radial distance where the Keplerian velocity is high \citep[e.g.,][]{Bir12,Lai12,Dra16}. Finally, sublimation and re-condensation of particles, that is, recycling of materials, around the snow line may trigger the solid pile-ups \citep[e.g.,][]{Ste88,Cie06,Bir10,Ros13,Mor15,Est16,Arm16,Sch17,Dra17,Ida16,Ida21,Hyo19,Hyo21b}.

The ND mechanism differs from the above mechanisms in that it does not require a pressure bump, snow line, and pebble growth and/or fragmentation. Instead, it requires an increase in $\alpha_{\rm mid}/\alpha_{\rm acc}$ with radial distance combined to the backreaction of solids to the gas that slows down the radial velocity of pebbles. The ND mechanism may be favored during the early phase of the disk evolution when the solid mass budget is large enough (i.e., a large $F_{\rm p/g}$) \citep{Dra21} and when an efficient gas accretion may take place with a dead zone (i.e., a small $\alpha_{\rm mid}/\alpha_{\rm acc}$) \citep{Lee21}. Our results highlight the importance of better observational constraints of gas and diffusion structure to understand whether the ND mechanism is indeed responsible for the dust trapping and formation of planetesimals (see also Sect.~\ref{sec_evolving_disk}).

\subsection{Application to evolving protoplanetary disks} \label{sec_evolving_disk}

In this study, using local simulations, $F_{\rm p/g}$ was set as a fixed constant at the outer boundary and the gas surface density was described by the classical $\alpha$-disk model ($\Sigma_{\rm g} \propto \alpha_{\rm acc}^{-1}$). This approach helped us to understand the nature and essence of the ND mechanism and the resultant planetesimal belt.

In reality, however, protoplanetary disks and $F_{\rm p/g}$ evolve as a function of time and the radial distance \citep[e.g.,][]{Dra21}. Discrete accretion episodes may exist \citep[e.g.,][]{Arm01,Zhu09} and $F_{\rm p/g}$ could be time-dependent. Indeed, \cite{Elb20} demonstrated short-period discrete accretion fluxes of pebbles at the early stages in the disk evolution \citep[see also][]{Ida21}. 

If the change in $F_{\rm p/g}$ is slow compared to the timescale to reach a steady width of the planetesimal belt (e.g., $\simeq$10\,kyr for $\tau_{\rm s}=0.1$ and $\simeq$100\,kyr for $\tau_{\rm s}=0.01$ at $r \lesssim 10$ au; Fig.\ref{fig_rate}), the outcomes would be similar to those observed in the late epochs in the simulations when planetesimals keep being formed within a fixed width (e.g., Fig.~\ref{fig_planetesmial}). In this case, the analytical estimations (e.g., Eq.~(\ref{eq_pla_ana})) can be used to predict the rate of planetesimal formation which does not depend on the radial structure of the dead zone and on $k_{\rm SI}$ (see Sect.~\ref{sec_edge}).

If it is the other way around, the width and profile of the planetesimal belt are time-dependent (e.g., when $F_{\rm p/g}$ changes within $\sim$10\,kyr for $\tau_{\rm s}=0.1$; Fig.\ref{fig_rate}). In this case, the rate of planetesimal formation depends on the disk radial structure (i.e., dead zone structure) and $k_{\rm SI}$. Such a case could be inferred by the lines in the early phase of the evolutions in Figs.~\ref{fig_planetesmial}-\ref{fig_rate} where the width is changing (see the lines in different colors in Figs.~\ref{fig_planetesmial}-\ref{fig_rate}). The exact timescale to reach a steady width of the planetesimal belt depends on the dead zone structure in the radial direction, $k_{\rm SI}$, and $\tau_{\rm s}$ (Figs.~\ref{fig_rate} and \ref{fig_tau}).

The values of the gas accretion parameter ($\alpha_{\rm acc}$) and of the parameter controlling the effect of vertical stirring of solid particles in the midplane ($\alpha_{\rm mid}$) are poorly known. Variations of the stellar accretion rate with age, the disk lifetime, and the extent of protoplanetary disks point to a disk evolution that should proceed with an effective viscosity parameter $\alpha_{\rm acc} \sim 10^{-3}-10^{-2}$ \citep{Hue05,Har16}. Separately, although it is an indirect estimation and thus one should be cautious interpreting data, dust settling compatible with ALMA observations seems to indicate a turbulence level of $\alpha_{\rm mid} \sim 0.1-1 \times 10^{-3}$ \citep{Mul12,Ove16,Pinte16}. Measurements of nonthermal gas motions point to similarly low turbulence values, a potential indication of $\alpha_{\rm mid} \sim 10^{-3}$ or less \citep{Bon16, Fla20}.

Our chosen parameters of $\alpha_{\rm acc}$ and $\alpha_{\rm mid}$ are within the range of the estimated values above. However, a deeper investigation of the disk structures, including the radial structure and $F_{\rm p/g}$, both from theoretical and observational sides are required to further test the applicability of the ND mechanism.

\subsection{Application to planet formation} \label{sec_planet_formation}

As discussed hereafter, planetesimal formation via the no-drift mechanism may prevent pebbles from drifting further inward and lead to the formation of seed planetesimals that would then grow by successive pebble accretion waves. 

First, in the ND scenario, pebbles drift inward from the outer disk region, entering a dead zone. As a consequence of the ND mechanism, the pebble mass flux that is propagated interior to the ND zone is reduced from $F_{\rm p/g}$ outside the dead zone to approximately $F_{\rm p/g,ND,min}$ inside the dead zone, as estimated from Eq.~(\ref{eq_Fpg_min}). This indicates the partial blocking of pebbles from flowing further inward. 

A stopping of drifting water-rich pebbles may be needed to avoid too much "wetting" of the inner planets, such as the terrestrial planets, to explain their observed water contents. The efficiency of pebble retention via the ND mechanism is proportional to the minimum $\alpha_{\rm mid}^{1/2}/\alpha_{\rm acc}$ value (Eq.~(\ref{eq_Fpg_min})). $F_{\rm p/g,ND,min}/F_{\rm p/g,given}$ is the fraction of pebbles that passes through the ND region. The other possible filtering mechanisms are the presence of pressure bumps, the growth of a large planetary core \citep{Gui14,Mor16,Bit21}, and/or the early consumption of solid materials in small size disks \citep{Ida19}.

Second, because the ND mechanism operates only while the criterion defined by  Eq.~\eqref{eq_ND_cri} is satisfied \citepalias[see also][]{Hyo21a}, the total mass of planetesimals depends on how long this condition is fulfilled (Sect.~\ref{sec_mass}). As discussed in Sect.~\ref{sec_evolving_disk}, the duration of the relatively high $F_{\rm p/g}$ for the ND mode to operate may be short. In this case, the total amount of planetesimals formed by the ND mechanism might be inadequate to fully form a giant planet (Fig.~\ref{fig_rate}). 

In this study, we assumed that pebbles whose Stokes number is $\tau_{\rm s} \sim 0.01 - 0.1$ so that pebbles are decoupled from the gas. Very small dust particles are well coupled with the gas and are thus not subject to the SI. Now, classically, pebbles are expected to be larger outside the snow line, due to more efficient sticking properties for ice than for rock \citep[e.g.,][]{Blu00}. This would imply that the ND mechanism with the SI may be favored at a greater radial distance than the snow line. Recent studies of material sticking properties, however, challenged this conventional view \citep[e.g.,][]{Kim15,Gun18,Mus19,Ste19}. In that case, rocky pebbles may also potentially experience the ND mode to form rocky planetesimals, if the criteria are satisfied (e.g., Fig.~\ref{fig_parameter} and Eq.~(\ref{eq_ND_cri})). 

Such considerations lead us to envision that the combination of planetesimal seeds formation via the ND mechanism with SI and successive pebble accretion \citep[e.g.,][]{Orm10,Lam14,Liu19} could lead to the formation of a large planetary core, even if the ND mode itself forms only a limited amount of planetesimals. We leave consideration of the pebble/planetesimal-to-planet formation in the context of the ND mechanism for future study.

Lastly, even if particles are too small for the SI to operate \citep[e.g., $\tau_{\rm s} < 0.01$;][]{Kat17} but when the ND mode develops with small $\tau_{\rm s}$, planetesimals may be formed via GI of runaway piling-up small dust particles. Studying GI in the context of the ND mechanism for small dust particles is beyond the scope of this paper and we leave that for future work.

\section{Summary} \label{sec_summary}

The no-drift (ND) mechanism is a newly identified instability mode in which a runaway pile-up of pebbles results from a variation of the vertical turbulence in the disk \citep[][\citetalias{Hyo21a}]{Hyo21a}. This process requires (1) the pebble-gas backreaction that slows down the radial velocity of pebbles and (2) a disk structure whose $\alpha_{\rm mid}/\alpha_{\rm acc}$ decreases with decreasing $r$, where $\alpha_{\rm acc}$ and $\alpha_{\rm mid}$ independently characterize the gas accretion (i.e., the surface density of gas) and the vertical turbulence (i.e., the scale height of pebbles), respectively. Such a nonuniform turbulent disk structure is motivated by recent nonideal magneto-hydrodynamics (MHD) simulations of a dead zone and a disk wind. Contrary to other planetesimal formation mechanisms, the ND mechanism does not require a pressure bump, snow line, and pebble growth and/or fragmentation.

In this study, we numerically modeled that pebbles drift inward from the outer region of a dead zone ($r \lesssim 10$ au), where the scale height of pebbles becomes smaller with decreasing $r$ (i.e., $\alpha_{\rm mid}$ decreases with decreasing $r$). The gas surface density was independently controlled by $\alpha_{\rm acc}$ in a classical $\alpha$-disk model.

Once the ND mode develops, the pile-up of pebbles continues in a runaway fashion for a fixed $F_{\rm p/g}$ without any consideration of planetesimal formation from accumulating pebbles \citepalias{Hyo21a}. In reality, however, planetesimals would form via the streaming instability as the midplane pebble-to-gas ratio of the spatial density becomes larger than unity via the ND mode. 

Here, we additionally included a prescription of the streaming instability in our 1D numerical code assuming a fixed $F_{\rm p/g}$ and using a modeled dead zone to study the resultant planetesimal belt formed by the ND mechanism and SI. The main findings are as follows:

\begin{itemize}
\setlength{\parskip}{0cm} 
\setlength{\itemsep}{0.2cm}

 \item[--] After pebbles enter a ND mode, planetesimals start to form via SI initially within a narrow ring whose width expands as accumulating pebbles radially diffuse over time (i.e., the radial width of the ND mode expands). The system finally reaches a steady-state in $\rho_{\rm p}/\rho_{\rm g}$, where the width of the planetesimal belt no longer changes, followed by continuous formation of planetesimals within a fixed width for a given fixed $F_{\rm p/g}$ and disk structure (Sect.~\ref{sec_overall} and Fig.~\ref{fig_planetesmial}).

\item[--] Planetesimal formation by the ND mechanism results from a complex interplay between a pile-up of pebbles drifting inward more slowly due to the backreaction effect, an absorption of inward-drifting pebbles, a radial diffusion of piled-up pebbles, and the conversion from pebbles to planetesimals via SI (Eq.~\eqref{eq_planetesimal}). Thus, the width and profile of the planetesimal belt depend on the disk structure and $k_{\rm SI}$ value. The smoother radial change in $\alpha_{\rm mid}/\alpha_{\rm acc}$ and/or smaller $k_{\rm SI}$ result in more radially-extended planetesimal belts (Figs.~\ref{fig_planetesmial} and \ref{fig_kSI}). Correspondingly, the timescale to reach a fixed width becomes longer (Fig.~\ref{fig_rate}) and so does for a smaller $\tau_{\rm s}$ (Fig.~\ref{fig_tau}).
 
\item[--] Once the planetesimal formation reaches a steady-state in $\rho_{\rm p}/\rho_{\rm g}$ with a fixed radial width for a given constant $F_{\rm p/g}$, the rate of planetesimal formation within the belt does not depend on the disk's radial structure and $k_{\rm SI}$, but it is regulated by $F_{\rm p/g}$ and the minimum value of $\alpha_{\rm mid}/\alpha_{\rm acc}$ (Sect.~\ref{sec_mass}, Fig.~\ref{fig_rate}, and Eq.~(\ref{eq_pla_ana})).

\item[--] For our nominal settings ($\dot{M}_{\rm g} \gtrsim 10^{-8}\,{\Msun}$\,yr$^{-1}$, $F_{\rm p/g} \gtrsim 0.1$, $\tau_{\rm s} \simeq 0.01-0.1$, $\alpha_{\rm mid} \lesssim 10^{-4}$, and $\alpha_{\rm acc} \simeq 10^{-3}-10^{-2}$ at $r\lesssim10$\,au with $k_{\rm SI} \gtrsim 10^{-3}$), a non-negligible total mass of planetesimals (more than Earth mass) is formed within about $10-100$\,kyr via the no-drift mode.
 
\end{itemize}

This study focused on the physics of the no-drift mechanism and the nature of the resultant planetesimal belt. To fully address the fate of a specific system where the ND is possibly occurring, we need full global simulations that include pebble growth and the disk viscous evolutions, and that self-consistently solve the dead zone structure. Such a completely self-consistently coupled numerical approach is conceptually and numerically challenging at this moment which will be the subject of future work.

\begin{acknowledgements}
R.H. acknowledges the financial support of JSPS Grants-in-Aid (JP17J01269, 18K13600). R.H. also acknowledges JAXA's International Top Young program. S.I. acknowledges the financial support (JSPS Kakenhi 15H02065), MEXT Kakenhi 18H05438). T.G. thanks the University of Tokyo and ELSI for their hospitality during 2019-2020 and acknowledges support from JSPS (Long-term fellowship L19506).  We thank the anonymous referee for thought-provoking comments which substantially improved the presentation of this manuscript.
\end{acknowledgements}

\bibliography{planetesimals}

\end{document}